\documentclass[referee]{aa520} 

\usepackage[dvips]{graphicx}
\usepackage{epsfig}
\usepackage{txfonts}

\newcommand{\grad}{\vec{\nabla}\,}

\renewcommand{\div}{\mathrm{div}\,}

\begin{document}

\graphicspath{{/home/petri/postdoc/rapport/images/}
  {/home/petri/postdoc/rapport/figure/}
  {/home/petri/postdoc/seminaires/} 
  {/home/petri/simulation/TN/image/}
  {/home/petri/simulation/TN/resultat/} 
  {/home/petri/simulation/TN/groupe/}
  {/home/petri/simulation/spectral/magnet/} 
  {/home/petri/simulation/spectral/hydro/}
  {/home/petri/simulation/spectral/hydro/lineaire/}
  {/home/petri/simulation/TN/photon/magnet/}
  {/home/petri/simulation/TN/photon/hydro/}
  {/home/petri/postdoc/articles/article1/}
  {/home/petri/postdoc/articles/article2/} 
  {/home/petri/postdoc/spectral/lineaire/hydro/}
  {/home/petri/simulation/spectral/hydro/} }

\title{Forced oscillations in a hydrodynamical accretion disk and QPOs.}

\author{J\'er\^ome P\'etri \inst{1}}

\offprints{J. P\'etri}

\institute {Astronomical Institute, University of Utrecht, P.O. Box
  80000, NL-3508 TA Utrecht, The Netherlands. 
  \thanks{\emph{Present address:} Max-Planck-Institut f\"ur Kernphysik,
    Saupfercheckweg 1, 69117 Heidelberg, Germany.}
}

\date{Received / Accepted}

\titlerunning{Forced oscillations in HD accretion disks}
\authorrunning{P\'etri}

\maketitle

\begin{abstract}
  
  This is the second of a series of papers aimed to look for an
  explanation on the generation of high frequency quasi-periodic
  oscillations~(QPOs) in accretion disks around neutron star, black
  hole, and white dwarf binaries. The model is inspired by the general
  idea of a resonance mechanism in the accretion disk oscillations as
  was already pointed out by Abramowicz \& Klu{\'z}niak
  (\cite{Abramowicz2001}). In a first paper
  (P\'etri~\cite{Petri2005a}, paper~I), we showed that a rotating
  misaligned magnetic field of a neutron star gives rise to some
  resonances close to the inner edge of the accretion disk. In this
  second paper, we suggest that this process does also exist for an
  asymmetry in the gravitational potential of the compact object. We
  prove that the same physics applies, at least in the linear stage of
  the response to the disturbance in the system.  This kind of
  asymmetry is well suited for neutron stars or white dwarfs
  possessing an inhomogeneous interior allowing for a deviation from a
  perfectly spherically symmetric gravitational field.  After a
  discussion on the magnitude of this deformation applied to neutron
  stars, we show by a linear analysis that the disk initially in a
  cylindrically symmetric stationary state is subject to three kinds
  of resonances: a corotation resonance, a Lindblad resonance due to a
  driven force and a parametric resonance.  In a second part, we focus
  on the linear response of a thin accretion disk in the 2D limit.
  Waves are launched at the aforementioned resonance positions and
  propagate in some permitted regions inside the disk, according to
  the dispersion relation obtained by a WKB analysis.  In a last part,
  these results are confirmed and extended via non linear
  hydrodynamical numerical simulations performed with a
  pseudo-spectral code solving Euler's equations in a 2D cylindrical
  coordinate frame. We found that for a weak potential perturbation,
  the Lindblad resonance is the only effective mechanism producing a
  significant density fluctuation.  In a last step, we replaced the
  Newtonian potential by the so called logarithmically modified
  pseudo-Newtonian potential in order to take into account some
  general-relativistic effects like the innermost stable circular
  orbit (ISCO). The latter potential is better suited to describe the
  close vicinity of a neutron star or a black hole.  However, from a
  qualitative point of view, the resonance conditions remain the same.
  The highest kHz QPOs are then interpreted as the orbital frequency
  of the disk at locations where the response to the resonances are
  maximal. It is also found that strong gravity is not required to
  excite the resonances.
  
\end{abstract}

\section{INTRODUCTION}

Accretion disks a very commonly encountered in the astrophysical
context. In the case where the accreting star is a compact object,
they offer a very efficient way to release the gravitational energy
into X-ray emission. However the process of release of angular
momentum leading to the accretion is still poorly understood. The
discovery of the high frequency quasi-periodic oscillations~(kHz-QPOs)
in the Low Mass X-ray Binaries (LMXBs) in~1996 offers a new tool for
the diagnostic of the physics in the innermost part of an accretion
disk and therefore in a strong gravitational field.

To date, quasi-periodic oscillations have been observed in about
twenty LMXBs sources containing an accreting neutron star. Among these
systems, the high-frequency QPOs (kHz-QPOs) which mainly show up by
pairs, possess strong similarities in their frequencies, ranging
from~300~Hz to about~1300~Hz, and in their shape~(see van der
Klis~\cite{vanderKlis2000} for a review).

Since this first discovery several models have been proposed to
explain the kHz-QPOs phenomenon observed in LMXBs. A beat-frequency
model was introduced to explain the commensurability between the twin
kHz-QPOs frequency difference and the neutron star rotation. This
interaction between the orbital motion and the star rotation happens
at some preferred radius. Alpar \& Shaham (\cite{Alpar1985}) and
Shaham~(\cite{Shaham1987}) proposed the magnetospheric radius to be
the preferred radius leading to the magnetospheric beat-frequency
model. The sonic-point beat-frequency model was suggested by Miller et
al.~(\cite{Miller1998}). In this model, the preferred radius is the
point where the radial inflow becomes supersonic. But soon after, some
new observations on Scorpius~X-1 showed that the frequency difference
is not constant~(van der Klis et al.~\cite{vanderKlis1997}). This was
then confirmed in other LMXBs like 4U~1636-53 (Jonker et
al.~\cite{Jonker2002}).  The sonic-point beat frequency model was then
modified to take into account this new fact~(Lamb \&
Miller~\cite{Lamb2001}).

The relativistic precession model introduced by Stella \& Vietri
(\cite{Stella1998},~\cite{Stella1999}) makes use of the motion of a
single particle in the Kerr-spacetime.  In this model, the kHz-QPOs
frequency difference is related to the relativistic periastron
precession of weakly elliptic orbits while the low-frequencies QPOs
are interpreted as a consequence of the Lense-Thirring precession.
Markovic \& Lamb (\cite{Markovic1998}) have also suggested this
precession in addition to some radiation warping torque which could
explain the low frequency QPOs.  More promisingly, Abramowicz \&
Klu{\'z}niak~(\cite{Abramowicz2001}) introduced a resonance between
orbital and epicyclic motion that can account for the 3:2 ratio around
Kerr black holes leading to an estimate of their mass and spin.
Indeed, for black hole candidates the 3:2 ratio was first noticed by
Abramowicz \& Klu{\'z}niak (\cite{Abramowicz2001}) who also recognized
and stressed its importance. Abramowicz et al.
(\cite{Abramowicz2003}) showed that the non-linear resonance for the
geodesic motion of a test particle can lead to the 3:2 ratio for the
two main resonances. Now the 3:2 ratio of black hole QPOs frequencies
is well established (McClintock \& Remillard, \cite{MacClintock2003}).
Klu{\'z}niak et al.~(\cite{Kluzniak2004a}) showed that the twin
kHz-QPOs can be explained by a non linear resonance in the epicyclic
motion of the accretion disk. The idea that a resonance in modes of
accretion disk oscillations may be excited by a coupling to the
neutron star spin is discussed by Klu{\'z}niak et
al.~(\cite{Kluzniak2004b}). Numerical simulations in which the disk
was disturbed by an external periodic field confirmed this point of
view (Lee et al.~\cite{Lee2004}). Rebusco~(\cite{Rebusco2004})
developed the analytical treatment of these oscillations.  More
recently, T\"or\"ok et al.~(\cite{Torok2005}) applied this resonance
to determine the spin of some microquasars. In other models, the QPOs
are identified with gravity or pressure oscillation modes in the
accretion disk (Titarchuk et al.~\cite{Titarchuk1998}, Wagoner et
al.~\cite{Wagoner2001}). Rezzolla et al. (\cite{Rezzolla2003})
suggested that the high frequency QPOs in black hole binaries are
related to p-mode oscillations in a non Keplerian torus.

Nevertheless, the propagation of the emitted photons in curved
spacetime can also produce some intrinsic peaks in the Fourier
spectrum of the light curves (Schnittman \& Bertschinger
\cite{Schnittman2004}). Bursa et al.~(\cite{Bursa2004}) suggested a
gravitational lens effect exerting a modulation of the flux intensity
induced by the vertical oscillations of the disk while simultaneously
oscillating radially. The propagation in the curved spacetime
reproduces also the 3:2 ratio observed in black hole binaries as shown
by Schnittman \& Rezzolla (\cite{Schnittman2005}).

Recent observations in accretion disks orbiting around white dwarfs,
neutron stars or black holes have shown a strong correlation between
their low and high frequencies QPOs (Mauche~\cite{Mauche2002}, Psaltis
et al.~\cite{Psaltis1999}). The relation is found to be the same for
any kind of compact object. This correlation has been explained in
terms of the centrifugal barrier model of Titarchuk et
al.~(\cite{Titarchuk2002}).

The very good agreement in the correlation of these low and high
frequencies QPOs spanned over more than 6~order of magnitude leads us
to the conclusion that the physical mechanism responsible for the
oscillations should be the same for the neutron star systems, the
black hole candidates and the cataclysmic variables (Warner et
al.~\cite{Warner2003}). Indeed, the presence or the absence of a solid
surface, a magnetic field or an event horizon play no relevant role in
the production of the X-ray
variability~(Wijnands~\cite{Wijnands2001}). In this paper we propose a
new resonance mechanism related to the evolution of the accretion disk
in a non axisymmetric rotating gravitational field.

This kind of forced response induced in the disk has been mostly
studied in the protoplanetary system or to explain rings around some
planets like Saturn. In the former case, a planet evolving within the
disk is responsible for the gravitational perturbation and should be
treated in the framework of hydrodynamical equations
(Tanaka~\cite{Tanaka2002}). In binary systems, the companion exerts
some torque on the accretion disk due to tidal forces. The spiral
pattern excited at the Lindblad resonance propagates down to the inner
boundary of the disk (Papaloizou \& Pringle \cite{Papaloizou1977},
Blondin \cite{Blondin2000}).  Whereas in the later case, this role is
devoted to the satellite and is well described (at least in first
approximation) by non collisional equations of motion as for instance
for the famous Saturn rings (Lissauer \& Cuzzi \cite{Lissauer1982}).
This simplified study helps to understand the physics of the resonance
without any complication introduced by the gaseous pressure (or the
radiation pressure) acting as a restoring force.

The paper is organized as follows. In Sec.~\ref{sec:EquInit}, we
describe the initial stationary state of the accretion disk and the
nature of the gravitational potential perturbation, starting with a
quadrupolar field to easily bring out the physics of the resonances
and, then, generalizing to a gravitational field possessing several
azimuthal modes in his Fourier transform. In Sec.~\ref{sec:AnalLin},
the governing equation for the linear regime of the Lagrangian
displacement is derived. Next, in Sec.~\ref{sec:DisqueMince}, we show
that the disk resonates due to the non axisymmetric component of the
gravitational potential. We study in detail the linear response of a
thin accretion disk which suffers no warping. Then a simplified three
dimensional analysis is carried out in Sec.~\ref{sec:AnaLinSimp}.
Finally, in Sec.~\ref{sec:Simulation}, in order to study the evolution
of the resonances on a longer timescale and in order to take into
account all the non-linearities, we perform 2D numerical simulations
by using a pseudo-spectral method which is compared to the linear
results.  First we apply this code to an accretion disk evolving in a
Newtonian potential. The results are then extended to the calculation
of a disk orbiting around a Kerr black hole by introducing a specific
pseudo-Newtonian potential. We briefly discuss the results in
Sec.~\ref{sec:Discussion}.  The conclusions of this work and the
possible generalization are presented in Sec.~\ref{sec:Conclusion}.

\section{THE INITIAL CONDITIONS}
\label{sec:EquInit}

In this section, we describe the initial hydrodynamical stationary
configuration of the accretion disk evolving in a perfectly
spherically symmetric gravitational potential created by the compact
object. We then give some justifications for the origin of the
gravitational perturbation.

\subsection{The stationary state}

In the equilibrium state, the disk possesses a stationary axisymmetric
configuration evolving in a spherically symmetric gravitational field
generated by the central star.  By axisymmetric we mean that every
field is invariant under rotation along the symmetry axis
$\partial/\partial\varphi=0$.  The disk inner and outer edges are
labeled by~$R_1$ and~$R_2$ respectively. All quantities possess only
a~$(r,z)$ dependence such that density~$\rho$, pressure~$p$ and
velocity~$\vec{v}$ are given by~:
\begin{equation}
  \label{eq:GrandeurEq}
  \rho = \rho(r,z), \;\; p = p(r,z), \;\; \vec{v} = r\,\Omega(r,z)\,\vec{e}_\varphi
\end{equation}
We can for instance assume that the disk is locally in isothermal
equilibrium and therefore uncouple the vertical $z$-structure from the
radial $r$-structure (Pringle \cite{Pringle1981}). The gravitational
attraction from the compact object is balanced by the centrifugal
force and the pressure gradient in the radial direction while in the
vertical direction we simply have the hydrostatic equilibrium. This
gives~:
\begin{eqnarray}
  \label{eq:Equilibre3D}
  \rho \, g_\mathrm{r} - \frac{\partial p}{\partial r} & = & -\rho \, \frac{v_\varphi^2}{r} \\
  \frac{\partial p}{\partial z} & = & \rho \, g_\mathrm{z}
\end{eqnarray}
We need only to prescribe the initial density in the disk. Assuming a
thin accretion disk, the gradient pressure will be negligible so that
the motion remains close to the Keplerian rotation. When a rotating
asymmetric component is added to the gravitational field, the
equilibrium state will be disturbed and evolves to a new configuration
where some resonances arise on some preferred radii which will be
determined in Sec.~\ref{sec:AnalLin}.

\subsection{Gravitational potential of a rotating star}
 
In the case of a rotating neutron star or white dwarf, the centrifugal
force induces a deformation of its shape and breaks the spherical
symmetry. The magnitude of this deformation depends on the equation of
state adopted for the star. 

Another reason for assuming a non spherical star is given by the
deformation due to the strong magnetic stress existing in the neutron
star's interior~(Bonazzola \& Gourgoulhon~\cite{Bonazzola1996}).  The
effect is very small with an ellipticity of the order of~$10^{-3}$
to~$10^{-6}$.  If the magnetic axis is not aligned with the rotation
axis of the star, the accretion disk will feel an asymmetric
gravitational field rotating at the same speed as the compact object.

We assume that the star is a perfect source of energy which means that
its spin rate remains constant in time.  To this approximation, energy
and angular momentum exchanges between star and disk has no influence
on the compact object. This assumption will be used throughout the
paper.

We insist on the fact that the goal of this paper is not to give an
accurate description of the origin and the shape of the deformation
but only to study the consequences of such a perturbation on the
evolution of the accretion disk.

Let's have a look on the shape of the potential induced by this
deformation of the stellar crust. To the lowest order, in an
appropriate coordinate frame attached to the star, the first
contribution is quadrupolar, there is no dipolar component.

The tensor of the quadrupolar moment~$D_{ij}$ can be reduced to a
traceless diagonal tensor in an appropriate coordinate
system~$(x,y,z)$ corresponding to the principal axis of the ellipsoid
formed by the star. In this particular coordinate frame we have
$D_{ij}=0$ for $i\ne j$ and $D_\mathrm{xx} + D_\mathrm{yy} +
D_\mathrm{zz} = 0$. The perturbed Newtonian potential expressed in
cylindrical coordinates~$(r,\varphi,z)$ is then given by~:
\begin{equation}
  \Phi(r,\psi,z) = - \frac{G\,M}{\sqrt{r^2+z^2}} \, \left( 1 + 
    \frac{r^2\, \cos^2\psi \, D_\mathrm{xx} + r^2\, \sin^2\psi \, D_\mathrm{yy} + 
      z^2 \, D_\mathrm{zz}}{2\,M\,(r^2+z^2)^2} \right) 
\end{equation}
This expression is only valid in the frame corotating with the star.
Returning back to an inertial frame, i.e. the observer frame, energy
and angular momentum are no longer conserved. Indeed, part of the
rotational energy of the star will be injected in the motion of the
accretion disk. This is the source of energy for the resonance to be
maintained. 

The physical relevant quantity is the perturbation in the
gravitational field caused by the quadrupolar moment and in the frame
corotating with the star, these components are given by~:
\begin{eqnarray}
\label{eq:GraviteQuad}
  \delta g_\mathrm{r} & = & - \, \frac{G\,M\,r}{(r^2+z^2)^{3/2}} \, 
  \frac{5 \, ( r^2 \, D_\mathrm{xx} \, \cos^2\psi + r^2 \, D_\mathrm{yy} \, \sin^2\psi
    + z^2 \, D_\mathrm{zz} ) - 2\, (r^2+z^2) \, ( D_\mathrm{xx} \,\cos^2\psi + 
    D_\mathrm{yy} \, \sin^2\psi) }{2\,M\,(r^2+z^2)^2} \nonumber \\
 & & \\ 
  \delta g_\varphi & = & \frac{G\,M\,r}{(r^2+z^2)^{3/2}} \, 
  \frac{(D_\mathrm{yy} - D_\mathrm{xx}) \, \cos\psi\,\sin\psi}{M\,(r^2+z^2)} \\
  \delta g_\mathrm{z} & = & - \, \frac{G\,M\,z}{(r^2+z^2)^{3/2}} \, 
  \frac{5 \, ( r^2 \, D_\mathrm{xx} \, \cos^2\psi + r^2 \, D_\mathrm{yy} \, \sin^2\psi
    + z^2 \, D_\mathrm{zz} ) - 2\, (r^2+z^2) \, D_\mathrm{zz}}{2\,M\,(r^2+z^2)^2}
\end{eqnarray}
To get the expression valid for a distant observer, at rest, we need
to replace~$\psi$ by~$(\varphi-\Omega_*\,t)$ where we have introduced
the rotation rate of the star by~$\Omega_*$. This potential seems far
from any realistic perturbation around a neutron star or a white
dwarf. Nevertheless, it offers a well understandable insight into the
resonances mechanisms by selecting solely a particular azimuthal
mode~$m$, namely the~$m=2$ quadrupolar case here. This kind of
gravitational perturbation is easily extended to more general shapes
including any other mode~$m$. The way to introduce naturally such a
structure is explained in the next subsection.

\subsection{Distorted stellar Newtonian gravitational field}
\label{sec:PotentielBoiteux}

The potential described above is a simple estimate of a quadrupolar
distortion induced by the rotation of the star. It introduce only one
azimuthal mode, allowing for a tractable analytical linear analysis.
Nevertheless, a more realistic view of the stellar field would include
several modes. These components can be introduced naturally in the
following way. We assume that the stellar interior is inhomogeneous
and anisotropic. In some regions inside the star, there exist clumps
of matter which locally generate a stronger or weaker gravitational
potential than the average. In order to compute analytically such kind
of gravitational field, we idealize this situation by assuming that
the star is made of homogeneous and isotropic matter everywhere (with
total mass~$M_*$).  To this perfect spherical geometry, we add a small
mass point~$M_\mathrm{p} \ll M_*$ located within the surface at a
position~$(R_\mathrm{p} \le R_*, \theta_\mathrm{p},\varphi_\mathrm{p}
= \Omega_*\,t)$. A finite size inhomogeneity can then be thought as a
linear superposition of such point masses.

Using spherical coordinates, the total gravitational potential induced
by this rotating star is~:
\begin{equation}
  \label{eq:PotBoiteux}
  \Phi(R,\theta,\varphi,t) = - \frac{G\,M_*}{|| \vec{R} ||} - 
  \frac{G\,M_\mathrm{p}}{|| \vec{R} - \vec{R_\mathrm{p}} ||}
\end{equation}
where the first term in the right hand side corresponds to the
unperturbed spherically symmetric gravitational potential whereas the
second term is induced by the small point like inhomogeneity.  For
simplicity, in the remainder of this paper, we suppose that the
perturber~$M_\mathrm{p}$ is located in the equatorial plane of the
star~$\theta_\mathrm{p} = \pi/2$.  The potential therefore becomes~:
\begin{equation}
  \label{eq:PotBoiteuxSimpl}
  \Phi(R,\theta, \psi) = - \frac{G\,M_*}{R} - 
  \frac{G\,M_\mathrm{p}}{\sqrt{R^2 + R_\mathrm{p}^2 - 2 \, R \, R_\mathrm{p} \, \cos\psi }}
\end{equation}
where the azimuth in the corotating frame
is~$\psi=\varphi-\Omega_*\,t$.  These potential can be Fourier
decomposed by using the Laplace coefficients~$b_{1/2}^m(x)$ as
follows~:
\begin{eqnarray}
  \label{eq:PotFourier}
  \Phi(R,\theta,\psi) & = & - \frac{G\,M_*}{R} - 
  \frac{G\,M_\mathrm{p}}{R} \, \sum_{m=0}^{+\infty} b_{1/2}^m\left( \frac{R_\mathrm{p}}{R} \right)
  \cos(m\,\psi) \\
  \label{eq:CoeffLaplace}
  b_{1/2}^m(x) & = & \frac{2-\delta_0^m}{2\,\pi} \, \int_0^{2\,\pi} 
  \frac{ \cos(m\,\psi)}{1 + x^2 - 2\,x\,\cos\psi} \, d\psi
\end{eqnarray}
where $\delta_0^m$ represents the Kronecker symbol.  The total linear
response of the disk is then the sum of each perturbation
corresponding to one particular mode~$m$. It is a generalization of
the quadrupolar field introduced in the previous section.

\begin{figure}[htbp]
  \centering
    \includegraphics[scale=1]{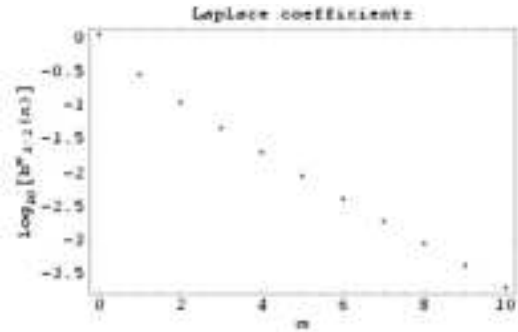}
  \caption{Laplace coefficients for the perturbed potential
    Eq.~(\ref{eq:PotBoiteuxCyl}) for~$x=0.5$.  Values are plotted on a
    logarithmic scale ($\log_{10}[b_{1/2}^m(x)]$) for $m\in[0..10]$.}
  \label{fig:CoeffLaplace}
\end{figure}

Having in mind to applied this results to a thin accretion disk, it is
preferable to use cylindrical coordinates such as~:
\begin{eqnarray}
  \label{eq:PotBoiteuxCyl}
  \Phi(r,\varphi,z,t) & = & - \frac{G\,M_*}{\sqrt{r^2+z^2}} - 
  \frac{G\,M_\mathrm{p}}{\sqrt{r^2 + r_\mathrm{p}^2 - 2 \, r \, r_\mathrm{p} \, \cos\psi + (z-z_\mathrm{p})^2 }} \\
  & = & - \frac{G\,M_*}{\sqrt{r^2+z^2}} - 
  \frac{G\,M_\mathrm{p}}{\sqrt{r^2+z^2}} \, 
  \sum_{m=0}^{+\infty} b_{1/2}^m\left( \frac{r_\mathrm{p}}{\sqrt{r^2+z^2}} \right)
  \cos(m\,\psi) 
\end{eqnarray}
The second expression as been obtained by assuming that the
inhomogeneity is located in the equatorial plane~$z_\mathrm{p}=0$.
The perturber is located inside the star and the disk never reaches
the stellar surface, therefore the Laplace coefficients~$b_{1/2}^m(x)$
never diverge because $x\le1$.  Moreover, because of the term
$\cos(m\,\psi)$ in the integrand Eq.~(\ref{eq:CoeffLaplace}), the
value of the Laplace coefficients decreases rapidly with the azimuthal
number~$m$. As a result, only the low azimuthal modes will influence
significantly the evolution of the disk. Keeping only the few first
terms in the expansion is sufficient to achieve reasonable accuracy.
An example of the numerical values of the Laplace
coefficients~$b_{1/2}^m(x)$ for~$x=0.5$ is shown in
Fig.~\ref{fig:CoeffLaplace}.

\section{LINEAR ANALYSIS}
\label{sec:AnalLin}

How will the disk react to the presence of this quadrupolar or
multipolar component in the gravitational field? To answer this
question, we can first study its linear response. To do this, we treat
each multipolar component as a small perturbation to the equilibrium
state prescribed in Sec.~\ref{sec:EquInit}. The hydrodynamical
equations of the accretion disk with adiabatic motion are given by~:
\begin{eqnarray}
  \label{eq:DiscHydroDens}    
  \frac{\partial\rho}{\partial t} + \div (\rho\,\vec{v}) & = & 0 \\
  \label{eq:DiscHydroVit}    
  \frac{\partial\vec{v}}{\partial t} + (\vec{v}\cdot\grad) 
  \, \vec{v} & = & \vec{g} - \frac{\grad p}{\rho} \\
  \label{eq:DiscHydroPres}    
  \frac{D}{Dt} \left( \frac{p}{\rho^\gamma} \right) & = & 0
\end{eqnarray}
All quantities have their usual meanings, $\rho$ being the density of
mass in the disk,~$\vec{v}$ the velocity of the disk, $p$ the gaseous
pressure, $\gamma$ the adiabatic index and~$\vec{g}$ the gravitational
field imposed by the star. Since we are not interested in the
accretion process itself, we neglect the viscosity whatever its origin
(molecular, turbulent, etc...).

\subsection{Lagrangian displacement}

Perturbing the equilibrium state with respect to the Lagrangian
displacement~$\vec{\xi}$, to first order we get for the Eulerian
perturbations of the density, velocity and pressure~:
\begin{eqnarray}
  \delta\rho & = & - \, \div(\rho\,\vec{\xi}) \\
  \delta\vec{v} & = & \frac{\partial\vec{\xi}}{\partial t} + 
  (\vec{v} \cdot \grad) \, \vec{\xi} - (\vec{\xi} \cdot \grad) \, \vec{v} \\
  \delta p & = & - \, \vec{\xi}\cdot\grad p - \gamma\,p\,\div\vec{\xi}
\end{eqnarray}
Making allowance for a perturbation in the gravitational field and
following the Frieman-Rotenberg analysis~(Frieman \&
Rotenberg~\cite{Frieman1960}), the Lagrangian displacement
satisfies a second order linear partial differential equation~:
\begin{eqnarray}
  \label{eq:PDEXi}
  \rho\,\frac{\partial^2\vec{\xi}}{\partial t^2} + 2\,\rho\,\vec{v}\cdot\vec{\nabla}\, 
  \frac{\partial\vec{\xi}}{\partial t} - \vec{\nabla}(\gamma\,p\,\div \vec{\xi}
  + \vec{\xi}\cdot\vec{\nabla}\,p) - 
  \div(\rho\,\vec{\xi}\,\vec{v}\cdot\vec{\nabla}\,\vec{v} - 
  \rho\,\vec{v}\,\vec{v}\cdot\vec{\nabla}\,\vec{\xi}) \nonumber & & \\ 
  + \vec{g} \, \div(\rho\,\vec{\xi}) + \div ( \rho \, \vec{\xi} ) \, \delta \vec{g} 
  & = & \rho \, \delta \vec{g}
\end{eqnarray}
We emphasize the fact that the last expression in the above equation
contains a term~$\div ( \rho \, \vec{\xi} ) \, \delta \vec{g}$ which
is of second order with respect to the perturbation and therefore
should be neglected. But in doing so, we suppress the parametric
resonance which will be studied in more detail below. Depending on
the magnitude of the perturbation, this instability will develop on a
timescale closely related to the amplitude of the perturbation and
should then not be ignored.

Introducing the convictive derivative by~$D/Dt = \partial_t +
\Omega\,\partial_\varphi$, we get for Eq.~(\ref{eq:PDEXi}) a more
concise form~:
\begin{equation}
  \label{eq:PDEXi2}
  \rho\,\frac{D^2\vec{\xi}}{Dt^2} - \vec{\nabla}(\gamma\,p\,\div \vec{\xi}
  + \vec{\xi}\cdot\vec{\nabla}\,p) - 
  \div(\rho\,\vec{\xi}\,\vec{v}\cdot\vec{\nabla}\,\vec{v} ) 
  + \vec{g} \, \div(\rho\,\vec{\xi}) +
  \div ( \rho \, \vec{\xi} ) \, \delta \vec{g} = \rho \, \delta \vec{g}
\end{equation}
Usually, when evolving in an axisymmetric gravitational field, the
above equation reduces to its traditional form
where~$\delta\vec{g}=\vec{0}$. However, in our treatment, due to the
gravitational perturbation, a driving force given by~$\rho \, \delta
\vec{g}$ appears. Moreover a parametric resonance is also involved due
to the term~$\div ( \rho \, \vec{\xi} ) \, \delta \vec{g}$. This will
be explained in the next section. 

Finding an analytical stability criteria for this system is
complicated or even an impossible task.  Furthermore, we cannot apply
the classical development in plane wave solutions leading to an
eigenvalue problem.  Indeed, the presence of some coefficients varying
periodically in time including~$\delta \vec{g}$ prevent from such a
treatment. However, the problem can be cast into a more convenient
form if we treat each component of the Lagrangian displacement as
independent variables and set the other two components equal to zero.
This simplified 3D analysis is done in section \ref{sec:AnaLinSimp}.
But before, we make a complete 2D linear analysis of
Eq.~(\ref{eq:PDEXi2}) in the following section.

\section{THIN DISK APPROXIMATION}
\label{sec:DisqueMince}

If we neglect the warping of the accretion disk, we can carry out a
more complete 2D linear analysis. Indeed, for a thin accretion disk,
its height~$H$ is negligible compared to its radius~$R$,~$(H/R)
\approx (c_\mathrm{s}/R\,\Omega) \ll 1$. We can give a detailed
analysis of the response of the disk to a linear perturbation by
setting~$\xi_\mathrm{z} = 0$ in the Eq.~(\ref{eq:PDEXi}) or
Eq.~(\ref{eq:PDEXi2}).

We seek solutions by writing each {\it perturbation}, such as the
components of the Lagrangian displacement~$\vec{\xi}$, those of the
{\it perturbed} velocity~$\delta\vec{v}$, the {\it perturbed}
density~$\delta \rho$, and the {\it perturbed} pressure~$\delta p$ as
\begin{equation}
  \label{eq:DvlptX}
  X(r,\varphi,t) = {\rm Re} [\tilde{X}(r) \, e^{i(m\,\varphi-\sigma\,t)}],
\end{equation}
where~$m$ is the azimuthal wavenumber and $\sigma$ the eigenfrequency
of the perturbation related to the speed pattern~$\Omega_\mathrm{p}$
by $\sigma = m\,\Omega_\mathrm{p}$.

Introducing the new unknown~$\psi=\sqrt{r\,p}\,\xi_\mathrm{r}$, it can
be shown that the problem reduces to the solution of a Schr\"odinger
type equation, (see Appendix~\ref{sec:AppDerSystProp})~:
\begin{equation}
  \label{eq:PsiSimp}
  \psi''(r) + V(r) \, \psi(r) = F(r)
\end{equation}
Eq.~(\ref{eq:PsiSimp}) is the fundamental equation we have to solve to
find the solutions to our problem far from the corotation resonance.
We refer the reader to Appendix~B of (P\'etri~\cite{Petri2005a},
paper~I) where we give an analytical method to find approximate
solutions to this equation.

The solutions of Eq.~(\ref{eq:PsiSimp}) divide into two classes of
different nature. The first one corresponds to free wave solutions
propagating in the accretion disk an associated with the homogeneous
part, $F(r)=0$. This gives rise to an eigenvalue problem in which the
pattern speed of the perturbation is determined by the specific
boundary conditions. The second one consists of a non-wavelike
disturbance associated with the inhomogeneous part, $F(r)\ne0$ due to
the gravitational perturbation. Here, the pattern speed of the density
perturbation is known and equal to the compact object rotation rate.
Therefore, there is no eigenvalue problem at this stage, we need only
to solve an usual ordinary differential equation with prescribed
initial conditions. 

\begin{figure}[htbp]
  \centering
  \begin{tabular}{cc}
  \includegraphics[scale=0.6]{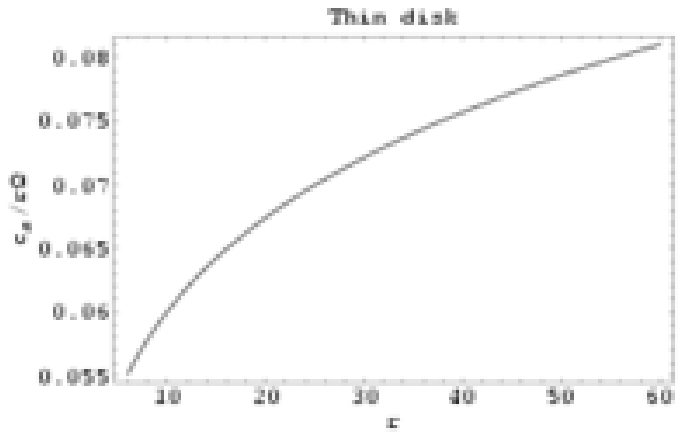} &
  \includegraphics[scale=0.6]{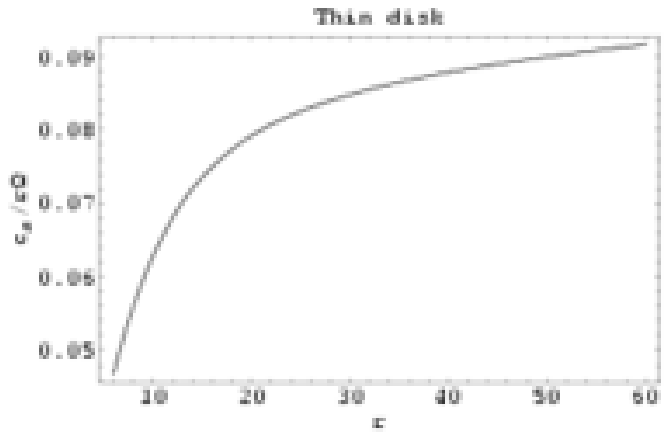} 
  \end{tabular}
  \caption{Thin disk approximation in the Newtonian case, on the left 
    and in the Schwarzschild case, on the right.  The
    ratio~$H/R=c_\mathrm{s}/R\,\Omega$ is plotted and remains less
    than~$1/10$ in the whole disk.}
  \label{fig:Epaisseur}
\end{figure}

For the purpose of numerical applications, the density profile in the
disk has the form~$\rho_0(r)=\frac{10^{-6}}{r}$. The adiabatic
constant is equal to~$\gamma=5/3$. The validity of the thin disk
approximation in the Newtonian and in the Schwarzschild case is
verified by plotting the ratio~$H/R$ as shown in
Fig.~\ref{fig:Epaisseur}. We now discuss them in more details in the
next subsections.

\subsection{Free wave solutions}

We compute the free wave solutions in order to show the influence of
the location of the inner boundary of the disk. When the disk
approaches the ISCO, the eigenfrequency of the waves increases.  Let's
start with a rough estimate. Looking for free wave solutions, a crude
estimate for the radial dependence is given by the WKB expansion as
follows~:
\begin{equation}
  \label{eq:OndeLibre}
    \Psi(r) = \Phi(r) \, e^{i\int^r k(s) \, ds}
\end{equation}
Putting this approximation into Eq.~(\ref{eq:PsiSimp}), the dispersion
relation is given by~:
\begin{equation}
  \label{eq:Dispersion}
  \omega^2 = \kappa^2 + c_\mathrm{s}^2 \, k^2
\end{equation}
Free waves can only propagate in regions where~$\omega^2 -
\kappa^2=V(r)\,c_\mathrm{s}^2\ge0$. The frontier between propagating
and damping zone is defined by the inner and outer Lindblad
radius~$r_\mathrm{L}^{in/out}$ defined by~$V(r_\mathrm{L}^{in/out}) =
0$.  Using the results of Appendix~B of paper~I, we can guess a better
approximation to the solution of the homogeneous
Eq.~(\ref{eq:PsiSimp}) which is valid even for~$r\approx
r_\mathrm{L}^{in/out}$. For the inner Lindblad resonance which is of
interest here, we introduce the following function~$\omega_1$,
writing~$r_\mathrm{L}=r_\mathrm{L}^{in}$~:
\begin{eqnarray}
  \omega_1(r) & = & - \left[ - \frac{3}{2} \, \int_{r_\mathrm{L}}^r \sqrt{V(s)} \, ds \right]^{2/3}
  \mathrm{for\;} r \le r_\mathrm{L} \\
  \omega_1(r) & = & \left[ \frac{3}{2} \, \int_{r_\mathrm{L}}^r \sqrt{-V(s)} \, ds \right]^{2/3}
  \mathrm{for\;} r \ge r_\mathrm{L}
\end{eqnarray}
The function~$\psi$ is then a linear combination of the 2~linearly
independent solutions~:
\begin{eqnarray}
  \label{eq:SolXi2}
  \psi_1(r) & = & \frac{Ai(\omega_1(r))}{\sqrt{|\omega_1'(r)|}} \\
  \psi_2(r) & = & \frac{Bi(\omega_1(r))}{\sqrt{|\omega_1'(r)|}}
\end{eqnarray}
Furthermore, we impose the solution to remain bounded as one boundary
condition, which leads to~$C_2=0$. Thus the solution for the
Lagrangian displacement is~: $\xi_\mathrm{r} =
C_1\,\psi_1(r)/\sqrt{r\,p}$.  At the inner boundary of the accretion
disk, the Lagrangian pressure perturbation should vanish. This is
expressed as~$\Delta p=0$ or equivalently~$\div\vec{\xi}=0$. To the
lowest order consistent with our approximation, the Lagrangian radial
displacement~$\xi_\mathrm{r}$ must satisfy~:
\begin{equation}
  \label{eq:Boundary}
  \xi'(R_1) + \left( 1 + 2\,m\,\frac{\Omega}{\omega} \right) \, \frac{\xi_\mathrm{r}(R_1)}{R_1} = 0
\end{equation}
This last condition determines the eigenfrequencies~$\sigma$ as a
function of the azimuthal mode~$m$. For any~$m$, there is an infinite
set of eigenvalues. However, the corresponding eigenfunctions become
more and more oscillatory, implying larger and larger wavenumber. In
the numerical applications, we shall restrict our attention to the ten
first one corresponding also to the highest values~$\sigma$.

The eigenvalues for the density waves are shown with decreasing value
in Table~\ref{tab:Eigenvalue}. This holds for a neutron star with
angular velocity $\nu_*=\Omega_*/2\pi=100$~Hz. We compared the
Newtonian case with the Schwarzschild metric. The highest speed
pattern given by~$\sigma/m$ never exceeds the orbital frequency at the
ISCO.

\begin{table}[htbp]
  \caption{The ten first highest eigenvalues~$\sigma$ for 
    the free wave solutions of Eq.~(\ref{eq:PsiSimp}).
    Values are normalized to the frequency of the ISCO,~$\Omega_\mathrm{ISCO}=6^{-3/2}$.
    Results are given for 3~azimuthal modes~$m=2,5,10$ as well for 
    the Newtonian as for the Schwarzschild gravitational field.}
  \label{tab:Eigenvalue}
  \centering
  \begin{tabular}{c c c | c c c}
    \hline
    \hline
    \multicolumn{6}{c}{Eigenvalues $\sigma/\Omega_\mathrm{ISCO}$} \\
    \hline
    \multicolumn{3}{c|}{Newtonian} & \multicolumn{3}{c}{Schwarzschild} \\
    \hline
    $m=2$ & $m=5$ & $m=10$ & $m=2$ & $m=5$ & $m=10$ \\
    \hline
    \hline
    0.838519 & 3.55997 & 8.22185 & 1.34337  & 4.01528 & 8.62713 \\
    0.60303  & 2.9742  & 7.26843 & 0.916075 & 3.2938  & 7.56642 \\
    0.468333 & 2.6023  & 6.6388  & 0.688728 & 2.84633 & 6.87224 \\
    0.373567 & 2.32075 & 6.15182 & 0.53807  & 2.51832 & 6.34571 \\
    0.302154 & 2.09279 & 5.74874 & 0.42896  & 2.25799 & 5.91539 \\
    0.246424 & 1.90117 & 5.40309 & 0.346084 & 2.04223 & 5.5494  \\ 
    0.20198  & 1.73634 & 5.09951 & 0.281241 & 1.85881 & 5.22998 \\
    0.166046 & 1.592   & 4.82837 & 0.229963 & 1.69932 & 4.94587 \\
    0.136682 & 1.46451 & 4.58301 & 0.188738 & 1.55983 & 4.68884 \\
    0.112755 & 1.35037 & 4.36067 & 0.153789 & 1.43488 & 4.45707 \\
    \hline
  \end{tabular}
\end{table}

\begin{figure}[htbp]
  \centering
  \begin{tabular}{cc}
  \includegraphics[scale=0.6]{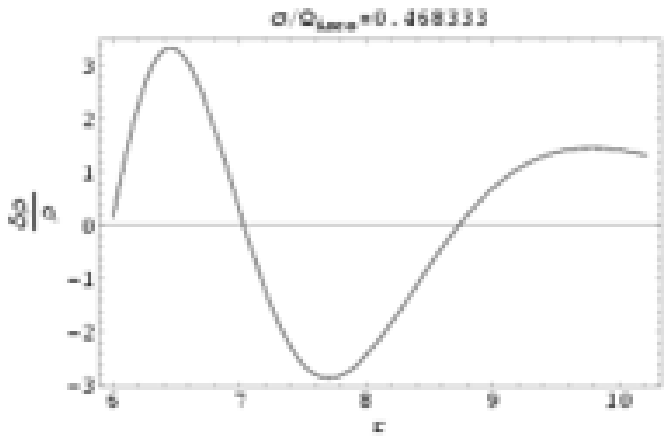} &
  \includegraphics[scale=0.6]{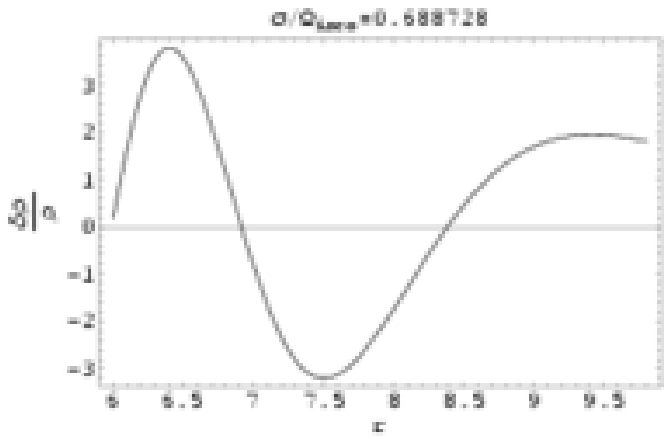} \\
  \includegraphics[scale=0.6]{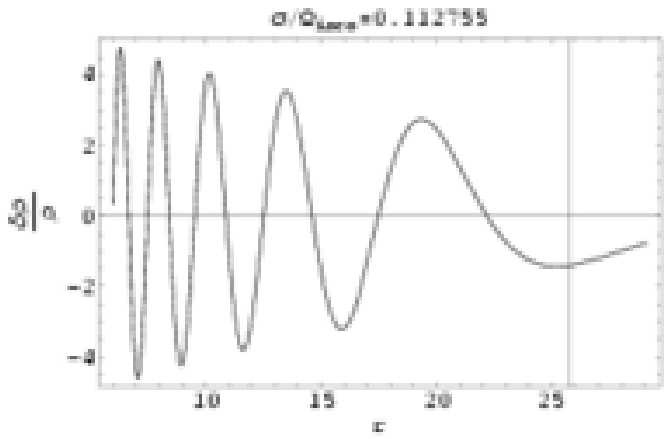} &
  \includegraphics[scale=0.6]{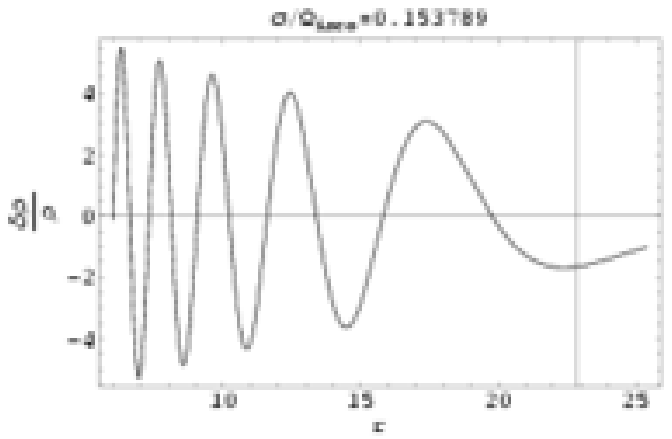}
  \end{tabular}
  \caption{Density wave perturbation in the disk caused by 
    the free wave propagation for the azimuthal mode~$m=2$.  Some
    examples are shown for different eigenvalues and for the Newtonian
    geometry, on the left, as well as for Schwarzschild one, on the
    right.  The vertical bar indicates the location of the inner
    Lindblad resonance. The normalization of the eigenfunctions is
    arbitrary.}
  \label{fig:FnPropre}
\end{figure}

Some examples of the corresponding eigenfunctions for the density
waves are shown in Fig.~\ref{fig:FnPropre} with arbitrary
normalization. Each of them possesses its own inner Lindblad radius
depending on the eigenvalue.

\begin{figure}[htbp]
  \centering
  \includegraphics[scale=0.6]{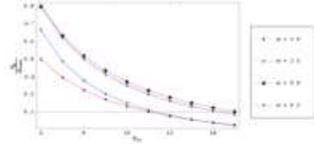}
  \caption{Variation of the highest eigenvalue, corresponding to 
    the eigenfunction having no node, as a function of the location of
    the inner edge of the disk~$R_\mathrm{in}$. There is a monotonic
    increase as the disk approaches the ISCO at~$R_\mathrm{in} = 6 \,
    R_\mathrm{g}$. Results are shown for the~$m=2,5$ modes in the
    Newtonian (N) and Schwarzschild (S) spacetime, respectively red
    and blue curves. The gravitational radius is defined
    by~$R_\mathrm{g} = G\,M_*/c^2$.}
  \label{fig:FnPropreRad}
\end{figure}

In a real accretion disk, the precise location of its inner edge does
not necessarily reach the ISCO, but can fluctuate due to the varying
accretion rate. For instance when the accretion process is enhanced,
the inner edge moves closer to the ISCO. As a results the highest
eigenvalue of the free waves also increases, see
Fig.~\ref{fig:FnPropreRad}.  When the ISCO is reached, the eigenvalues
does not change anymore because the boundary conditions remains at the
ISCO and the eigenfrequencies saturate.  This kind of saturation of
the QPO frequency has been observed in some LMXB as reported for
instance in a paper by Zhang et al.~(\cite{Zhang1998}).  The accretion
disk has probably reached its ISCO in this particular system. Relating
the free wave solutions to this QPO cut off mechanism is not obvious
at this stage of our study. Indeed, exciting the waves with a
frequency (the star rotation rate~$\Omega_*$) different from its
eigenfrequency ($\sigma$) would require a non-linear process not taken
into account in our model so far.  However, we will show that already
in the linear stage, for sufficiently strong amplitude in the
perturbation field, the parametric resonance explains some interesting
features of the kHz QPOs (see Sec.~\ref{sec:Parametric}).

\subsection{Driven wave perturbations}
\label{sec:Nonwave}

These driven waves are useful to check our numerical scheme described
in Sec.~\ref{sec:Simulation}. Indeed, in the non-linear simulation,
the free wave solutions decay and only the forced solution will
survive on a very long timescale.  We now solve the full inhomogeneous
Eq.~(\ref{eq:PsiSimp}) to seek for the solution corresponding to the
non-wavelike perturbation in the case of a quadrupolar field
perturbation.  The quadrupolar momentum of the gravitational field due
to the non-spherical rotating star is given by~:
\begin{eqnarray}
  \delta g_\mathrm{r}(r,\varphi,t) & = & \frac{3}{4} \, \frac{G}{r^4} \, \left( 
  D_\mathrm{xx} + D_\mathrm{yy} + (D_\mathrm{xx} - D_\mathrm{yy} ) \, e^{2\,i\,(\varphi - \Omega_*\,t)} \right) \\
  \delta g_\varphi(r,\varphi,t) & = & \frac{i}{2} \, \frac{G}{r^4} \, (D_\mathrm{xx} - D_\mathrm{yy} )
  \, e^{2\,i\,(\varphi - \Omega_*\,t)}
\end{eqnarray}
In the numerical applications, we choose~$D_\mathrm{xx}$
and~$D_\mathrm{yy}$ such that~$D_\mathrm{xx}+D_\mathrm{yy}$ remains
negligible with respect to~$D_\mathrm{xx}-D_\mathrm{yy}$. In the
complex amplitudes of~$\delta g_\mathrm{r/\varphi}$ we therefore only
keep the radial dependence for the mode~$m=2$. Thus~:
\begin{eqnarray}
  \delta g_\mathrm{r}(r) & = & \frac{3}{4} \, \frac{G}{r^4} \, (D_\mathrm{xx} - D_\mathrm{yy} ) \\
  \delta g_\varphi(r) & = & \frac{i}{2} \, \frac{G}{r^4} \, (D_\mathrm{xx} - D_\mathrm{yy} )
\end{eqnarray}
We have to solve the second order ordinary differential equation
for~$\psi$ with the appropriate boundary conditions
Eq.~(\ref{eq:Boundary}). The solution is~:
\begin{equation}
  \label{eq:PsiPart}
  \psi_\mathrm{r}(r) = C_1 \, \psi_1(r) + C_2 \, \psi_2(r) +
  \pi \, \mathrm{sign}(\omega_1'(r)) \, \int_{r_\mathrm{L}}^r \left( \psi_1(r) \, \psi_2(s) 
    - \psi_1(s) \, \psi_2(r) \right) F(s) \, ds
\end{equation}
The constant~$C_2$ is chosen such that the solution remains bounded
for~$r\gg r_\mathrm{L}$~:
\begin{equation}
  \label{eq:C2}
  C_2 = \lim_{r\to\infty} \pi \, \mathrm{sign}(\omega_1') \, \int_{r_\mathrm{L}}^r
  \psi_1(s) \, F(s) \, ds
\end{equation}
This integral is convergent because the function~$\psi_1$ is
exponentially decreasing with the radius.
\begin{figure}[htbp]
  \centering
  \begin{tabular}{cc}
    \includegraphics[scale=0.6]{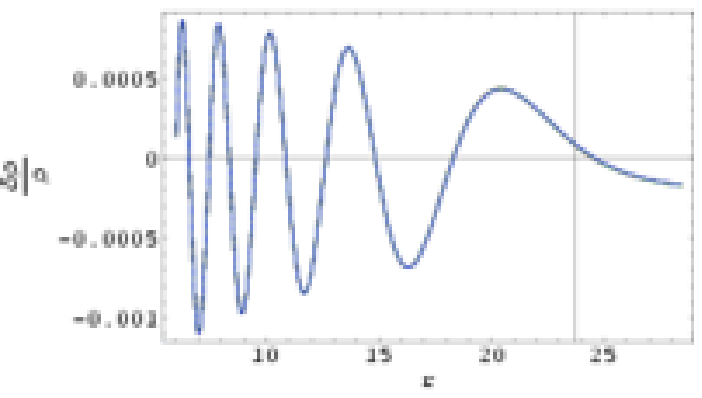} &
    \includegraphics[scale=0.6]{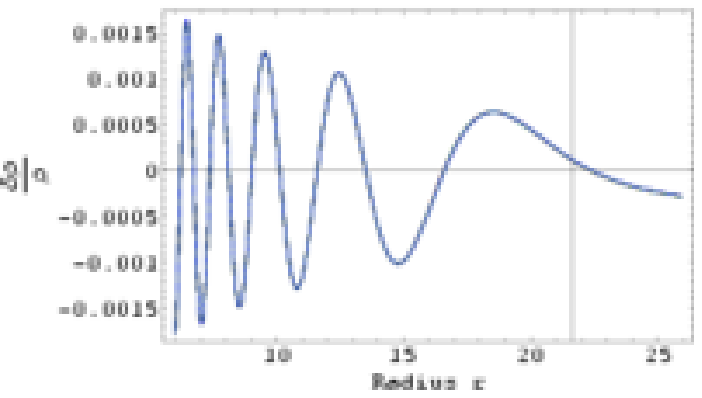}
  \end{tabular} 
  \caption{Non-wavelike density perturbation in a Newtonian potential, 
    on the left, and in a Schwarzschild potential, on the right, for
    the mode~$m=2$ and the speed pattern~$\sigma=2\,\Omega_*$. The
    amplitude of the disturbances is related to the strength of the
    gravitational perturbation and is therefore not arbitrary.}
  \label{fig:SolForcee}
\end{figure}
The analytical solutions Eq.~(\ref{eq:PsiPart}) agree well with the
direct numerical integration of Eq.~(\ref{eq:PsiSimp}). On the
Fig.~\ref{fig:SolForcee}, there is no graphical distinction between
them nor for the Newtonian potential neither for the Schwarzschild
field.

\subsection{Corotation resonance}

The corotation resonance is defined by the radius~$r_\mathrm{c}$
where~$\omega_*(r_\mathrm{c}) = 0$. Actually, this equation possesses
two solutions corresponding to~$\omega(r_\mathrm{c}) = \pm
\frac{m\,c_\mathrm{s}(r_\mathrm{c})}{r_\mathrm{c}}$.  The width of
this region is of the order of the disk height~$O(H)$.  For very thin
disks, this discrepancy can be neglected and the two solutions merge
together in an unique corotation radius given by~$\omega(r_\mathrm{c})
= 0$.  In other words, we assume in this case that~$\omega=\omega_*$.
However, in our numerical application, the separation between the two
corotation radii is large enough to be resolved. For the detailed
study of both corotation, we have to use the more accurate
Eq.~(\ref{eq:AppXiRCorot}).  Keeping only the leading divergent terms
in the coefficients of the ordinary differential equation, we obtain~:
\begin{eqnarray}
  \label{eq:XiRCorot2}
  \frac{m^2\,c_\mathrm{s}^4}{r^2\,\omega_*^2} \, \xi_\mathrm{r}''(r) + \frac{m^2\,c_\mathrm{s}^4}{r^2} \, 
  \frac{d}{dr}\left(\frac{1}{\omega_*^2}\right) \, \xi_\mathrm{r}'(r) +
  2\, \frac{m\,c_\mathrm{s}^2\,\Omega\,\omega}{r} \, 
  \frac{d}{dr}\left(\frac{1}{\omega_*^2}\right) \, \xi_\mathrm{r}(r)
  = i \frac{m\,c_\mathrm{s}^2}{r} \, \frac{d}{dr}
  \left( \frac{1}{\omega_*^2} \right) \, \delta g_\varphi(r) \nonumber \\
\end{eqnarray}
We introduce the new independent variable~:
\begin{equation}
  \label{eq:rc2x}
  x = \frac{r-r_\mathrm{c}}{r_\mathrm{c}}
\end{equation}
Developing~$\omega_*$ to the first order around the corotation
radius~$r_\mathrm{c}$ we have~:
\begin{equation}
  \omega_*(r) = \omega_*(r_\mathrm{c}) + (r-r_\mathrm{c})\,\omega_*'(r_\mathrm{c}) + o(r-r_\mathrm{c})
  = x\,r_\mathrm{c}\,\omega_*'(r_\mathrm{c}) + o(x) \approx \alpha \, x
\end{equation}
To this approximation, we have to solve~:
\begin{equation}
  \label{eq:diff}
  \xi_\mathrm{r}''(x) - \frac{2}{x} \, \xi_\mathrm{r}'(x) - 
  4 \, \frac{\Omega\,\omega\,r_\mathrm{c}^2}{m\,c_\mathrm{s}^2\,x} \, \xi_\mathrm{r}(x) =
  -2\,i\,\frac{r_\mathrm{c}^2\,\delta g_\varphi(r_\mathrm{c})}{m\,c_\mathrm{s}^2\,x} 
\end{equation}
This is of the form~:
\begin{equation}
  \label{eq:yODE}
  y''(x) - \frac{2}{x} \, y'(x) - \frac{b}{x} \, y(x) = \frac{c}{x}
\end{equation}
with~$b=4 \, \frac{\Omega\,\omega\,r_\mathrm{c}^2}{m\,c_\mathrm{s}^2\,x}\ge0$
and~$c=-2\,i\,\frac{r_\mathrm{c}^2\,\delta g_\varphi(r_\mathrm{c})}{m\,c_\mathrm{s}^2}$.  Making
the change of variable~$t=2\,\sqrt{b\,x}$ and introducing the new
unknown~$v(t)$ by~$y(t)=t^3\,v(t)$, it can be shown that~$v(t)$
satisfies the modified Bessel equation of order~3~:
\begin{equation}
  \label{eq:BesselModif}
  v''(t) + \frac{1}{t} \, v'(t) - ( 1 + \frac{9}{t^2} ) \, v(t) = 0
\end{equation}
This is solved by~:
\begin{equation}
  \label{eq:BesselModifSol}
  v(t) = C_1 \, I_{3}(t) + C_2 \, K_{3}(t)
\end{equation}
Thus, the complete most general solution to Eq.~(\ref{eq:diff}) for
which a particular solution is easily found to be a constant equal
to~$\xi_\mathrm{r}^p(r) = i \, \frac{\delta g_\varphi(r_\mathrm{c})}
{2\,\Omega\,\omega}$, is
\begin{equation}
  \label{eq:n}
  \xi_\mathrm{r}(x) = C_1 \, x^{3/2} \, I_3(2\,\sqrt{b\,x}) + 
  C_2 \, x^{3/2} \, K_3(2\,\sqrt{b\,x}) +
  i \, \frac{\delta g_\varphi(r_\mathrm{c})}{2\,\Omega\,\omega}
\end{equation}
Finally, near the corotation radius, the Lagrangian displacement which
remains bounded needs~$C_1=0$~:
\begin{equation}
  \label{eq:XiRForce}
  \xi_\mathrm{r}(x) = C_2 \, x^{3/2} \, K_3(2\,\sqrt{b\,x}) + 
  i \, \frac{\delta g_\varphi(r_\mathrm{c})}{2\,\Omega\,\omega}
\end{equation}

The density disturbance induced in the disk by the rotating
gravitational perturbation is then to the lowest order~:
\begin{equation}
  \label{eq:drhorho}
  \frac{\delta\rho}{\rho} = - \frac{\div(\rho\vec{\xi})}{\rho} = 
  - \frac{1}{\rho\,r}\,\frac{d}{dr}(r\,\rho\,\xi_\mathrm{r})
  + i\,\frac{m}{r\,\omega_*^2} \, \left( \delta g_\varphi +
    2\,i\,\Omega\,\omega\,\xi_\mathrm{r} \right)
\end{equation}
The displacement~Eq.~(\ref{eq:XiRForce}) is continuous and
differentiable everywhere. Thus, the first term on the right hand side
has a finite value. The second term on the RHS needs a special
treatment.  Indeed, when~$r$ approaches~$r_\mathrm{c}$ the numerator and the
denominator vanish as well, leaving us with an undetermined expression
of the form~$0/0$. To find the behavior near~$r_\mathrm{c}$ we note that near
the corotation, $(\delta g_\varphi + 2\,i\, \Omega\, \omega\,
\xi_\mathrm{r})$ behaves as $(\delta g_\varphi(r)-\delta
g_\varphi(r_\mathrm{c})) = \delta g_\varphi'(r_\mathrm{c}) \,
(r-r_\mathrm{c})$ with~$\delta g_\varphi'(r_\mathrm{c})\ne0$.  Thus we
conclude that it approaches zero as~$x$ and
\begin{equation}
  \label{eq:drhorhoapprox}
  \frac{\delta\rho}{\rho} \approx i \, \frac{m\,\delta g_\varphi'(r_\mathrm{c})}
  {\omega_*'(r_\mathrm{c})^2 \, r_\mathrm{c}^2 \, x}
\end{equation}
The divergent term in the density perturbation
Eq.~(\ref{eq:drhorhoapprox}) tends to infinity as~$\frac{1}{x}$. This
result is consistent with the conclusions drawn by Goldreich \&
Tremaine~(\cite{Goldreich1979}) for a disk without self-gravity.

From the comparison between Newtonian and Schwarzschild geometry, we
conclude that the introduction of general relativistic effects as the
ISCO does not changes the qualitative behavior of the disk
response. The ISCO only shifts the location of the Lindblad
resonances and the eigenvalues of the free wave solutions.

\subsection{Counterrotating disk}

\begin{figure}[htbp]
  \centering
  \includegraphics[scale=0.8]{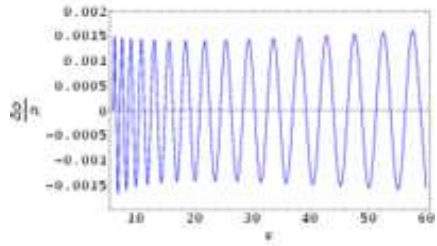}
  \caption{Density perturbation~$\delta\rho/\rho$
    in the counterrotating disk evolving in a Newtonian potential.
    Results are given for the mode~$m=2$ and the
    eigenvalue~$\sigma=-2\,\Omega_*$.}
  \label{fig:SolForceeContre}
\end{figure}

We can also use the previous analysis for a counterrotating accretion
disk. Because the spin of the star does not intervene in the
homogeneous Schr\"odinger equation, the free wave solutions remain
identical to the above mentioned results. The only change comes from
the non-wavelike disturbance for which we have to
replace~$\Omega_*\rightarrow-\Omega_*$. Following the same outline as
in~\ref{sec:Nonwave}, we solve numerically the inhomogeneous
Schr\"odinger equation and we also looked for an analytical
approximate solution. The results showing the density perturbation is
plotted in Fig.~\ref{fig:SolForceeContre}. The two solutions are
graphically indistinguishable, the discrepancy is less than~1~\%.  In
this special case, the Lindblad resonances do not appear anymore in
the computational domain.

\section{SIMPLIFIED ANALYSIS}
\label{sec:AnaLinSimp}

\subsection{The eigenvalue problem}

To get more insight in the nature of the resonances, we focus now only
on the displacement of the disk in each direction independently,
setting~$\xi_i=0$ in the other directions. This means that we neglect
the coupling between the oscillations occurring in perpendicular
directions. Despite this simplification, this will help us to bring
out the meaning of the oscillations and to derive some resonance
conditions without removing any physically meaningful mechanism.

Let's begin the study with the motion in the vertical direction,
setting~$(\xi_\mathrm{r}, \xi_\varphi) = 0$, we find~:
\begin{equation}
  \rho\,\frac{D^2\xi_\mathrm{z}}{Dt^2} - \frac{\partial}{\partial z}\left( \gamma\,p\,
    \frac{\partial\xi_\mathrm{z}}{\partial z} \right) - \rho \,
  \xi_\mathrm{z}\,\frac{\partial g_\mathrm{z}}{\partial z} + \frac{\partial}{\partial z} 
    \left( \rho \, \xi_\mathrm{z} \right) \, \delta g_\mathrm{z} = \rho \, \delta g_\mathrm{z}
\end{equation}
Developing the vertical component of the gravity near the equatorial
plane to the first order in~$z$, it can be cast into the
form~$g_\mathrm{z}(r,z)=-\kappa_\mathrm{z}^2(r)\,z$. The vertical epicyclic
frequency~$\kappa_\mathrm{z}$ depends only on the radius. So we get~:
\begin{equation}
  \label{eq:PDEXiz}
  \frac{D^2\xi_\mathrm{z}}{Dt^2} - \frac{1}{\rho} \, 
  \frac{\partial}{\partial z}\left( \rho\,c_\mathrm{s}^2 \,
    \frac{\partial\xi_\mathrm{z}}{\partial z} \right) + \kappa_\mathrm{z}^2\,\xi_\mathrm{z}
  + \frac{1}{\rho} \, \frac{\partial}{\partial z}\left( \rho \, \xi_\mathrm{z} \right) \, \delta g_\mathrm{z} 
  = \delta g_\mathrm{z}
\end{equation}
We have introduced the sound speed by~$c_\mathrm{s}^2 = \frac{\gamma\,p}{\rho}$.

The same can be done for the radial motion by setting~$(\xi_\mathrm{z},
\xi_\varphi) = 0$, we obtain a similar expression~:
\begin{equation}
  \label{eq:PDEXir}
  \frac{D^2\xi_\mathrm{r}}{Dt^2} - \frac{1}{\rho\,r} \, 
  \frac{\partial}{\partial r}\left( \rho \, c_\mathrm{s}^2 \, r \,
    \frac{\partial\xi_\mathrm{r}}{\partial r} \right) + \kappa_\mathrm{r}^2\, \xi_\mathrm{r} + \frac{1}{\rho\,r} \, 
  \frac{\partial}{\partial r}\left( r\,\rho\,\xi_\mathrm{r}\right) \, \delta g_\mathrm{r}
  = \delta g_\mathrm{r}
\end{equation}
The exact value of the radial epicyclic frequency depends on the
pressure distribution in the disk. 

Eq.~(\ref{eq:PDEXiz}) and~(\ref{eq:PDEXir}) look very similar, the
discrepancy coming only from the difference between the planar and the
cylindrical geometry (terms containing~$r$). We recognize in the two
first terms of Eq.~(\ref{eq:PDEXiz}) and~(\ref{eq:PDEXir}) a sound
wave propagation equation in a tube of spatial varying but time
independent cross section~(Morse \& Feshbach~\cite{Morse1953}). The
first and third term put together is an harmonic oscillator at the
epicyclic frequency~$\kappa_\mathrm{r/z}$. So the three first terms
are a generalization of the Klein-Gordon equation and do not give rise
to any kind of instability. The interesting part are those containing
the perturbation in the gravitational field~$\delta g_\mathrm{r/z}$.
Neglecting the sound wave propagation, we recognize a kind of Hill
equation corresponding to an oscillator with periodically time-varying
eigenfrequency. It is well known that this type of equation shows what
is called a parametric resonance.  Moreover due to the rotation of the
star, this perturbation will vary sinusoidally in time and the Hill
equation specializes to the Mathieu equation for which we know the
resonance conditions. Indeed, Mathieu equation written in the form
\begin{equation}
  \label{eq:Mathieu}
  y''(t) + \omega_0^2 \, ( 1 + h \, \cos\gamma\,t) \, y(t) = 0  
\end{equation}
becomes unstable if the excitation frequency~$\gamma =
2\frac{\omega_0}{n}$ where~$n$ is an integer (Landau \&
Lifshitz~\cite{Landau1982}). Note that the resonant frequency does not
depend on the amplitude~$h$. The corresponding growth rates are
proportional to~$h^n$. For small amplitudes of the excitation term,
only the first few integer~$n$, let's say~$n\le5$, are relevant for
this parametric instability.

The equation contains also an harmonic oscillator excited by a driven
force given by~$\delta g_\mathrm{r/z}$. This gives rise to the well
known driven resonance.

A careful analysis of Eq.~(\ref{eq:PDEXiz}) and Eq.~(\ref{eq:PDEXir})
shows that in the frame locally corotating with the disk, the
Lagrangian displacement feels a modulation due to the gravity
perturbation term~$\delta g_\mathrm{r/z}$. In this corotating frame,
its time dependence contains expressions like~$\cos m(\varphi -
(\Omega_* - \Omega)\,t)$ and~$\sin m(\varphi - (\Omega_* -
\Omega)\,t)$, (Eq.~\ref{eq:GraviteQuad}). Therefore the modulation
occurs at a frequency~$m\,|\Omega_* - \Omega|$. Each Fourier mode of
the perturbed gravitational field contributes to give its proper
modulation frequency.

From this analysis we expect three kind of resonances corresponding to~:
\begin{itemize}
\item a {\it corotation resonance} at the radius where the angular
  velocity of the disk equals the rotation speed of the star. This is
  only possible for prograde disks. The resonance condition to
  determine the corotating radius is simply~:
  \begin{equation}
    \label{eq:ResCorot}
    \Omega = \Omega_* 
  \end{equation}
\item a {\it inner and outer Lindblad resonances} at the radius where
  the radial or vertical epicyclic frequency equals the frequency of
  each mode of the gravitational potential as seen in the locally
  corotating frame. We find the resonance condition to be~:
  \begin{equation}
    \label{eq:ResForcage}
    m \, | \Omega_* - \Omega | = \kappa_\mathrm{r/z}
  \end{equation}
  The name for this resonance arises from the analogy with the theory
  of density waves met in the context of the spiral structure in
  galactic dynamics. If the pressure acts as a restoring force in the
  vertical direction, the derivation of the vertically driven
  resonances is given by $m \, | \Omega_* - \Omega | = \sqrt{1+\Gamma}
  \, \kappa_\mathrm{z}$, where $\Gamma=\partial\ln P/\partial\ln \rho$
  is the effective adiabatic index, (see Lubow~\cite{Lubow1981}). We
  will not go into such refinement for a first approach to the
  resonance problem.
\item a {\it parametric resonance} related to the time-varying
  epicyclic frequency, (Hill equation). The rotation of the star induces
  a sinusoidally variation of the epicyclic frequency leading to the well
  known Mathieu's equation. The resonance condition can be derived as followed~:
  \begin{equation}
    \label{eq:ResPara}
    m \, |\Omega_* - \Omega| = 2 \, \frac{\kappa_\mathrm{r/z}}{n}
  \end{equation}
\end{itemize}
Note that the driven resonance is a special case of the parametric
resonance for~$n=2$. However, their growth rate differ by the
timescale of the amplitude magnification. Driving causes a linear
growth in time while parametric resonance causes an exponential
growth.

The parametric resonance condition Eq.~(\ref{eq:ResPara}) has been
derived on the basis of a single particle orbit perturbation without
taking into account the fluid description of the gas. Hirose and
Osaki~(\cite{Hirose1990}) applied this method in the context of
tidally distorted accretion disk in cataclysmic variable. However,
Lubow~(\cite{Lubow1991}) showed that a more careful treatment of the
resonances in the hydrodynamical case leads to some eccentric
instability as well to unstable tilt of the accretion disk due to mode
coupling (Lubow~\cite{Lubow1992}). Due to the effect introduce by the
fact that the disk is a fluid, the growth rate of this instability is
only quadratic in the strength of the tidal force.  Nevertheless,
another parametric instability arising in a tidally distorted
accretion disk has been introduced by Goodman~(\cite{Goodman1993}). He
showed that the growth rate is linear in the tidal force amplitude and
propagates only in a three-dimensional disk.

\subsection{Results}
\label{sec:Parametric}

\subsubsection{Newtonian disk}
From the resonance conditions derived above, Eq.~(\ref{eq:ResForcage})
and Eq.~(\ref{eq:ResPara}), we can find the radii where each of this
resonance will occur. Beginning with the Newtonian potential, it is
well known that the angular velocity, the radial and epicyclic
frequencies for a single particle are all equal so that~$\Omega =
\kappa_\mathrm{r} = \kappa_\mathrm{z}$. This conclusion remains true for a thin
accretion disk having~$c_\mathrm{s}/R\,\Omega\ll1$.  Distinguishing between the
two signs of the absolute value, we get for the parametric resonance
condition Eq.~(\ref{eq:ResPara}), which includes the Lindblad
resonance for the special case~$n=2$, the following orbital rotation
rate~:
\begin{equation}
  \frac{\Omega}{\Omega_*} = \frac{m}{m \pm 2/n}  
\end{equation}
As a consequence, the resonances are all located in the frequency
range~$\Omega\in[\Omega_*/3, 3\,\Omega_*]$. 

In table~\ref{tab:ResPara}, we indicate the results for a~$300$~Hz and
a~$600$~Hz spinning neutron star and for the first three mode~$m$ and
for the integer~$n=1,2$. Because of the degeneracy~$\kappa_\mathrm{r} =
\kappa_\mathrm{z}$, the resonances in the radial and vertical directions occur
at exactly the same locations.

\begin{table}[h]
    \caption{Value of the orbital frequencies at the parametric
      resonance radii for the first three order~$n$ in the case of a
      Newtonian gravitational potential.  The results are given for
      a~$1.4\,\mathrm{M}_\odot$ neutron star rotating respectively at~$300$~Hz
      and~$600$~Hz. The value on the left of the symbol~/ corresponds
      to the absolute value sign taken to be~- and on the right to
      be~+.}
  \label{tab:ResPara}
  \begin{center}
    \begin{tabular}{c | c c | c c}
      \hline
      \hline
       Azimuthal mode~$m$ & \multicolumn{4}{c}{Orbital frequency $\nu(r,a)$ (Hz)} \\
      \hline
      & \multicolumn{2}{c|}{$\nu_*=600$~Hz} & \multicolumn{2}{c}{$\nu_*=300$~Hz} \\
      & $n=1$ & $n=2$ & $n=1$ & $n=2$ \\
      \hline
      \hline
      1 & -600 / 200 & ---- / 300 & -300 / 100 & --- / 150 \\
      2 & ---- / 300 & 1200 / 400 &  --- / 150 & 600 / 200 \\ 
      3 & 1800 / 360 &  900 / 450 &  900 / 180 & 450 / 225 \\
      \hline
    \end{tabular}
  \end{center}
\end{table}

The pair of highest orbital frequencies for the $\nu_*=300$~Hz
spinning neutron star are $\nu_1=900$~Hz and $\nu_2=600$~Hz. The peak
separation frequency is then~$\Delta\nu=300$~Hz~$=\nu_*$. The
vertical motion induced by the parametric resonance at that location
will appear as a modulation in the luminosity of the accretion disk.
For the 600~Hz spinning neutron star, the highest orbital frequencies
are 1800~Hz and 1200~Hz. However, due to the ISCO, the former one is
not observed because it is located inside the ISCO and therefore does
not correspond to a stable orbit. Therefore, the first two highest
observable frequencies are $\nu_1=1200$~Hz and $\nu_2=900$~Hz.
The peak separation frequency becomes then~$\Delta\nu=300$~Hz~$ =
\nu_*/2$. Thus the peak separation for slow spinning neutron stars is
$\Delta\nu=\nu_*$ whereas for fast spinning neutron star the peak
separation is $\Delta\nu=\nu_*/2$. This segregation between slow and
fast rotating neutron stars is well observed in several accreting
systems~(van der Klis \cite{vanderKlis2004}). These conclusions
confirm the results already obtained in the single particle
approximation (P\'etri \cite{Petri2005b}, \cite{Petri2005c}).

We believe that only oscillations in the vertical direction can give
rise to significant periodic changes in the accretion disk luminosity.
However, resonances associated with the radial epicyclic frequency are
shown because we start with a 2D study in the equatorial plane (linear
analysis and numerical simulations). Radial oscillations in the disk
are hardly observable because they will not lead to a significant
change in the luminosity as would be the case for a warped disk for
instance. The warping is induced at some preferred radii where the
resonance conditions for vertical oscillations are fulfilled. This
study would necessitate a full 3D treatment of the accretion disk
which is left for future work. Nevertheless, the properties of the
propagation of waves in a three-dimensional accretion disk have been
investigated by many authors. Lubow and Pringle~(\cite{Lubow1993})
studied the linear behavior of free waves in an isothermal disk.
Korycansky and Pringle~(\cite{Korycansky1995}) extended this work to
the case of polytropic disks and showed that the local 2D dispersion
relation is not valid anymore. Due to the stratified vertical
structure, waves are refracted and reach the surfaces of the disk~(Lin
et al.~\cite{Lin1990a}).  This happens within a distance of the order
of~$r_\mathrm{L}/m$ and is called wave-channeling by Lubow \&
Ogilvie~(\cite{Lubow1998}) who undertook a detailed study of the wave
propagation in the neighborhood of the Lindblad resonances.  Finally,
the 3D response to a tidal force was explored by Lin et
al.~(\cite{Lin1990b}). The behaviour of the waves launched at the
resonance radii (corotation, Lindblad and parametric) propagating in
the 3D disk is a complicated task. In this paper, we just start with a
simple picture, focusing on the resonances itself without taking into
account the propagation of the disturbances.

\subsubsection{General relativistic disk}

When the inner edge of the accretion disk reaches values of a few
gravitational radii, the general relativistic effects become
important. The degeneracy between the three
frequencies~$\Omega$,~$\kappa_\mathrm{r}$ and~$\kappa_\mathrm{z}$ will be removed and
they will depend on the angular momentum of the star~$a$.  Indeed we
distinguish 3~characteristic frequencies in the accretion disk around
a Kerr black hole (or a rotating neutron star)~:
\begin{itemize}
\item the orbital angular velocity~:
  \begin{equation}
    \label{eq:FreqOrbit}
    \Omega(r,a)=\frac{1}{r^{3/2}+a}    
  \end{equation}
\item the radial epicyclic frequency~:
  \begin{equation}
    \label{eq:FreqRadial}
    \kappa_\mathrm{r}(r,a)=\Omega(r,a)\,\sqrt{1-\frac{6}{r} + 8\,\frac{a}{r^{3/2}}
      - 3\,\frac{a^2}{r^2} }
  \end{equation}
\item the vertical epicyclic frequency~:
  \begin{equation}
    \label{eq:FreqVertical}
    \kappa_\mathrm{z}(r,a)=\Omega(r,a)\,\sqrt{1 -4\,\frac{a}{r^{3/2}} +
      3\,\frac{a^2}{r^2}}
  \end{equation}
\end{itemize}
The parameter~$a$ corresponds to the angular momentum of the star, in
geometrized units. We also assume that~$G\,M_*=1$. For a neutron star
of mass~$M_*$ and rotating at the angular velocity~$\Omega_*$, it is
given by $a=\frac{c\,I_*}{G\,M_*^2}\,\Omega_*$.

We have the following ordering~:
\begin{eqnarray}
\Omega>\kappa_\mathrm{z}>\kappa_\mathrm{r} & for & \;\; a>0 \\
\kappa_\mathrm{z}>\Omega>\kappa_\mathrm{r} & for & \;\; a<0
\end{eqnarray}

The parametric resonance conditions Eq.~(\ref{eq:ResPara}) splitted
into the two cases become~:
\begin{equation}
  \label{eq:ResParaGR}
  \Omega(r,a) \pm \frac{2\,\kappa_\mathrm{r/z}(r,a)}{m\,n} = \Omega_* 
\end{equation}

For a given angular momentum~$a$, we have to solve these equations for
the radius~$r$. For a neutron star, we adopt the typical parameters~:
\begin{itemize}
\item mass~$M_*=1.4\,\mathrm{M}_\odot$~;
\item angular velocity~$\nu_*=\Omega_*/2\pi=300-600$~Hz~;
\item moment of inertia~$I_*=10^{38}\;kg\,m^2$~;
\item angular momentum~$a_*=\frac{c\,I_*}{G\,M_*^2}\,\Omega_*$.
\end{itemize}
The angular momentum is then given by~$a_*=5.79*10^{-5}\,\Omega_*$.
Solving Eq.~(\ref{eq:ResParaGR}) for the radius and then deducing the
orbital frequency at this radius we get the results shown in
tables~\ref{tab:ResParaGRrad} and~\ref{tab:ResParaGRver}. For the spin
rate of the star we find~$a_*=0.1-0.2$ and so the vertical epicyclic
frequency remains close to the orbital
one~$\kappa_\mathrm{z}\approx\Omega$.  Thus for the vertical
resonance, we are still approximately in the Newtonian case mentioned
in the previous section and the same conclusions apply here to.
Consequently, the relativistic results are the same as those discussed
in the previous section dealing with a Newtonian disk. The only
difference comes from the presence of the ISCO added in a
self-consistent way by changing the behaviour of the radial epicyclic
frequency which vanishes at the inner edge.

\begin{table}[h]
  \caption{Value of the orbital frequencies at the radial parametric
    resonance radii for the first three order~$n$ in the general
    relativistic Kerr spacetime.  The results are given for
    a~$1.4\,\mathrm{M}_\odot$ neutron star rotating respectively at~$300$~Hz
    and~$600$~Hz. The value on the left of the symbol~/ corresponds
    to the absolute value sign taken to be~- and on the right to
    be~+.}
  \label{tab:ResParaGRrad}
  \centering
  \begin{tabular}{c | c c | c c}
    \hline
    \hline
    Azimuthal mode~$m$ & \multicolumn{4}{c}{Orbital frequency $\nu(r,a)$ (Hz)} \\
    \hline
    & \multicolumn{2}{c|}{$\nu_*=600$~Hz} & \multicolumn{2}{c}{$\nu_*=300$~Hz} \\
    & $n=1$ & $n=2$ & $n=1$ & $n=2$ \\
    \hline
    \hline
    1 & 2542 / 212 & 1566 / 319 &  1257 / 106 & 779 / 159 \\
    2 & 1566 / 318 &  955 / 419 &   779 / 159 & 477 / 210 \\ 
    3 & 1135 / 380 &  809 / 468 &   566 / 190 & 404 / 234 \\
    \hline
  \end{tabular}
\end{table}

\begin{table}[h]
  \caption{Value of the orbital frequencies at the vertical parametric
    resonance radii for the first three order~$n$ in the general
    relativistic Kerr spacetime.  The results are given for
    a~$1.4\,\mathrm{M}_\odot$ neutron star rotating respectively at~$300$
    and~$600$~Hz. The value on the left of the symbol~/ corresponds
    to the absolute value sign taken to be~- and on the right to
    be~+.}
  \label{tab:ResParaGRver}
  \begin{center}
    \begin{tabular}{c | c c | c c}
      \hline
      \hline
      Azimuthal mode~$m$ & \multicolumn{4}{c}{Orbital frequency $\nu(r,a)$ (Hz)} \\
      \hline
      & \multicolumn{2}{c|}{$\nu_*=600$~Hz} & \multicolumn{2}{c}{$\nu_*=300$~Hz} \\
      & $n=1$ & $n=2$ & $n=1$ & $n=2$ \\
      \hline
      \hline
      1 & ---- / 200 & ---- / 300 &  --- / 100 & --- / 150 \\
      2 & ---- / 300 & 1198 / 400 &  --- / 150 & 599 / 200 \\ 
      3 & 1790 / 360 &  899 / 450 &  898 / 180 & 450 / 225 \\
      \hline
    \end{tabular}
  \end{center}
\end{table}

  
We emphasize the fact that these results apply to a rotating
asymmetric magnetic field with exactly the same resonance conditions
Eq.~(\ref{eq:ResCorot})-(\ref{eq:ResForcage})-(\ref{eq:ResPara})
provided that the flow is not to far from its Keplerian motion, i.  e.
a weakly magnetized accretion disk with high~$\beta$-plasma parameter
where this parameter is defined by (Delcroix \& Bers
\cite{Delcroix1994}, see also paper~I)~:
\begin{equation}
  \label{eq:betaPlasma}
  \beta = \frac{p}{B^2/2\mu_0}
\end{equation}
$p$ being the pressure and $B$ the local magnetic field strength.

\section{NUMERICAL SIMULATIONS}
\label{sec:Simulation}

Now we have identified the resonance location (Lindblad, parametric
and corotation) in the disk due to the perturbation in the
gravitational field, we go further to include the full non linearities
of the hydrodynamical equations by performing 2D simulations in the
($r,\varphi$) plane. This is the goal of the next section.

\subsection{Linear analysis}

In order to check the numerical pseudo-spectral code, we solve the
full non-linear HD equations with a weak $m=2$ azimuthal perturbation.
We retrieve the results mentioned in section~\ref{sec:Nonwave}. This
is discussed in the following subsections. 

We use the geometrized units for which~$G=c=1$. The distances are
measured with respect to the gravitational radius given
by~$R_\mathrm{g}=G\,M_*/c^2$. Moreover, the simulations are performed
for a star with~$M_*=1$ so that in the new units we
have~$R_\mathrm{g}=1$.  In all the simulations presented below, the
star is assumed to be an ellipsoid with the main axis given
by~$R_\mathrm{x} = R_\mathrm{z} \ne R_\mathrm{y}$. The standard
resolution is~$N_\mathrm{R} \times N_\varphi=256\times32$ where
$N_\mathrm{R}$ and $N_\varphi$ are the number of grid points in the
radial and azimuthal direction respectively.

Before the time~$t=0$, the disk stays in its axisymmetric equilibrium
state and possesses only an azimuthal motion. At~$t=0$, we switch on
the perturbation by adding the quadrupolar component to the
gravitational field. We then let the system evolve during more than
one thousand orbital revolutions of the inner edge of the disk. We
performed four sets of simulations. In the first one, the
gravitational potential was Newtonian. In the second one, we used a
pseudo-Newtonian potential in order to take into account the ISCO.
This is well suited to describe the Schwarzschild spacetime. In the
third one, we took into account the angular momentum of the star by
introducing a pseudo-Kerr geometry. And finally in the fourth and last
set, we performed simulations with a counter rotating accretion disk
evolving in a Newtonian potential described in the first set. 

We perform the simulation in the thin disk limit. For this thin
gaseous disk, there is a slightly difference of the order~$(H/R)^2$
between the single particle characteristic frequencies and the true
disk frequencies where~$H$ is the typical height of the disk and~$R$
its radius. Indeed, inspecting Eq.~(\ref{eq:PDEXir}) and neglecting
the gravitational perturbation $\delta g_\mathrm{r}$, a rough
estimation in order of magnitude gives $\partial/\partial r \approx
R^{-1}$ and due to the thin disk approximation we also have
$c_\mathrm{s} \approx \Omega_\mathrm{k} \, H$ where
$\Omega_\mathrm{k}$ is the Keplerian orbital frequency for a single
particle. Therefore the coefficient in front of $\xi_\mathrm{r}$ is
approximately $\Omega_\mathrm{k}^2 \, ( 1 - H^2/R^2$), baring in mind
that $\kappa_\mathrm{r} \approx \Omega_\mathrm{k}$ and proving the
aforementioned statement. In all our simulations, we choose the
physical parameters such that this ratio remains less than~$1/10$,
spanning roughly from~$0.05$ to~$0.09$. In such a way the single
particle approximation remains valid within~10~\%. Typical behaviors
of the ratio~$H/R$ are depicted for the Newtonian and Schwarzschild
case in Fig.~\ref{fig:Epaisseur}. If the disk were assumed to be thick
so that $H\approx R$, the difference between single particle and fluid
frequency can be appreciable. Moreover, the orbital motion remains no
longer Keplerian in such a geometry.  We now deal with the results.

\subsection{Newtonian potential}
\label{sec:ResNewt}

First, we study the behavior of the thin disk in the Newtonian
potential.  In this case, the Keplerian rotation rate, the vertical
and the radial epicyclic frequencies for a single particle are all
identical as discussed before. To a good approximation we have~:
\begin{equation}
\Omega_\mathrm{k} \approx \kappa_\mathrm{r} \approx \kappa_\mathrm{z}
\end{equation}

The star normalized rotation rate around the z-axis is equal
to~$\Omega_*=0.0043311$. Assuming a~$1.4\,\mathrm{M}_\odot$ neutron
star, this corresponds to a spin of~$\nu_*=100.0$~Hz. The disk inner
boundary is located at~$R_1=6.0$ while the outer boundary is located
at~$R_2=60.0$.  The orbital angular motion at the inner edge of the
disk is~$\Omega_\mathrm{in}=R_1^{-3/2} = 0.0680$. We normalize the
time by dividing it by the spin period of the star~$T_* =
\frac{2\,\pi}{\Omega_*} = 1450.7$.

\begin{figure}[h]
  \begin{center}
    \includegraphics[draft=false,scale=0.6]{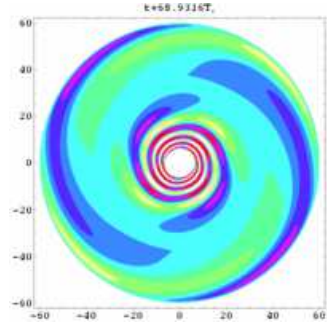}
    \caption{Final snapshot of the density perturbation~$\delta\rho/\rho_0$
      in the accretion disk evolving in a quadrupolar perturbed
      Newtonian potential.  The disk extends from~$R_1=6.0$
      to~$R_2=60.0$. The rotation rate of the star
      is~$\Omega_*=0.0043311$.  The time is normalized to the spin
      period~$T_*=1450.7$. The $m=2$ structure emerges in relation
      with the~$m=2$ quadrupolar potential perturbation.}
    \label{fig:DensDiscNewt} 
  \end{center}
\end{figure}

\begin{figure}[h]
  \begin{center}
    \includegraphics[draft=false,scale=0.8]{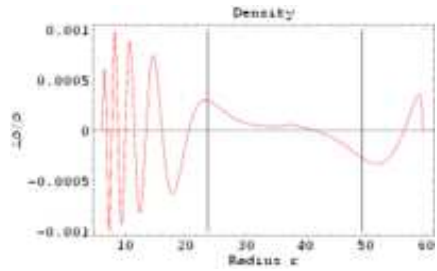}
    \caption{Cross section of the density perturbation~$\delta\rho/\rho_0$
      in the Newtonian disk at the final time of the simulation. The
      inner and outer Lindblad resonances appear clearly
      at~$r_\mathrm{L}^{in/out}=23.7/49.3$. When crossing the corotation
      resonance the density curve shows a break in its slope.}
    \label{fig:DensDiscNewtSect} 
  \end{center}
\end{figure}

The time evolution of the density perturbation in the disk calculated
as~$\Delta\rho/\rho_0=\rho/\rho_0-1$ is shown in
Fig.~\ref{fig:DensDiscNewt}. The corotation resonance located
at~$r=40.0$ expected from the condition Eq.~(\ref{eq:ResCorot}) is not
visible at this stage. Indeed the weak linear growth rate makes the
corotation resonance relevant only after a few~$10^5$ orbital
revolutions which is hundreds of time more than the time of the
simulation. However, after a few hundred of orbital revolutions, the
disk settles down to a new quasi-stationary state in which the inner
and outer Lindblad resonances persist. This happens after a transition
regime where density waves leave the disk by crossing its edges.
Almost all of the energy put into the disk by the star's rotation
leaves the computational domain at its inner and outer edges.  The non
reflecting boundary conditions act as a kind of viscosity strongly
damping the oscillations. We refer the reader to paper~I for the
method of implementation of these non reflecting boundary conditions.
This is confirmed by inspection of the Fig.~\ref{fig:DensDiscNewtSect}
in which we have plotted a cross section of the final density
perturbation~$\delta\rho/\rho_0$ for a given azimuthal angle,
namely~$\varphi=0$. As expected the density perturbation vanishes at
the disk edges. The shape of this perturbation agrees well with the
linear analysis. Indeed comparing Fig.~\ref{fig:DensDiscNewtSect} and
the left part of Fig.~\ref{fig:SolForcee}, the only small difference
comes from the different boundary conditions imposed. Nevertheless,
the location of the inner and outer Lindblad resonances as well as the
number of roots of each function and the maximum amplitude are equal.

\begin{figure}[h]
  \begin{center}
    \includegraphics[draft=false,scale=0.8]{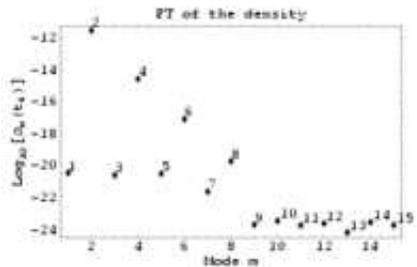}
    \caption{Amplitude of the Fourier components of the density perturbation. 
      $t_f$ stands for the final time of the numerical simulation.
      The axisymmetric mode is not represented. The odd modes are
      numerically zero. Due to the small nonlinearities, the even
      modes are apparent but with a weak amplitude. The
      components~$m\ge9$ are set to zero because of the desaliasing
      process.}
    \label{fig:DensTFDiscNewt} 
  \end{center}
\end{figure}

The nonlinearities are therefore weak for the whole simulation
duration.  Indeed, looking at the Fourier spectrum of the density in
Fig.~\ref{fig:DensTFDiscNewt}, where the amplitude of each component
is plotted vs the mode~$m$ on a logarithmic scale.  The odd modes are
not present.  However, the weak nonlinearities create a cascade to
high even modes starting with~$m=2$. The largest asymmetric expansion
coefficient~$C_m$ is~$m=2$, the next even coefficients follow roughly
a geometric series with a factor~$q=10^{-3}$, so we can write for
all~$m$ even, $C_m \approx q^{m/2-1} \, C_2$ until they reach values
less than~$10^{-20}$ which can be interpreted as zero from a numerical
point of view. The deviation from the stationary state being weak, the
amplitudes of these even modes decay compared to the previous one, the
highest being of course~$m=2$. Thus, even in the full non linear
simulation, the regime remains quasi linear. As a conclusion, the
parametric resonance phenomenon discussed in the previous section is
irrelevant at this stage of our work. The effect of strong
nonlinearity putting the system out of its linear regime will be
studied in a forthcoming paper. Note that due to the desaliasing
process, the modes~$m\ge9$ are all set to zero.  Note also that the
free wave solutions leave the computational domain and are no longer
present. Only the non-wavelike disturbance produces significant
changes in the density profile.

\begin{figure}[h]
  \begin{center}
    \includegraphics[draft=false,scale=0.6]{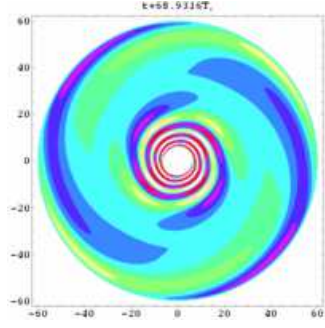}
    \caption{Same as Fig.~\ref{fig:DensDiscNewt} but for the higher
      $512\times64$~resolution. There is no difference between the two
      runs proving that all the structure in the disk is resolved.}
    \label{fig:DensDiscNewt2} 
  \end{center}
\end{figure}

\begin{figure}[h]
  \begin{center}
    \includegraphics[draft=false,scale=0.8]{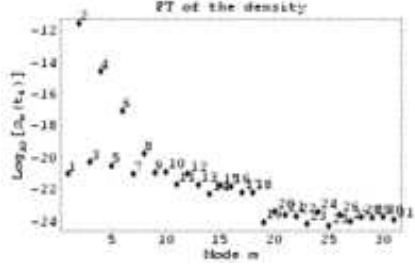}
    \caption{Same as Fig.~\ref{fig:DensTFDiscNewt} but for the
      $512\times64$~resolution. The components~$m\ge19$ are set to
      zero because of the desaliasing process. However, the
      coefficients are already close to zero for~$m\ge9$ without
      desaliasing.}
    \label{fig:DensTFDiscNewt2} 
  \end{center}
\end{figure}

The resolution of~$256\times32$ which seems rather low is nevertheless
justified by the fact that the components of the Fourier-Tchebyshev
transform decay rapidly and become negligible after the first few
terms in the Fourier expansion and after the first 30 or 40 terms in
the Tchebyshev expansion.  In order to check that this resolution is
however sufficient, we performed a new set simulation by increasing
the number of grid points by a factor two in both coordinates, namely
we used a resolution of~$512\times64$. The density perturbation is
then given by the Fig.~\ref{fig:DensDiscNewt2} and the Fourier
expansion coefficients in Fig.~\ref{fig:DensTFDiscNewt2} which have to
be compared respectively with Fig.~\ref{fig:DensDiscNewt} and
Fig.~\ref{fig:DensTFDiscNewt}. The desaliasing process does not affect
the Fourier coefficient because they vanish already well
before~$m=19$. This proves that the structure is spatially fully
resolved with both resolutions. Indeed, there is actually no
significant difference between the 2 pictures. The lowest resolution
reaches already a very good numerical precision. We therefore keep
this~$256\times32$ resolution in the next sections.

\subsection{Pseudo-Schwarzschild potential}

In order to take into account the modification of the radial epicyclic
frequency due to the curved spacetime around a Schwarzschild black
hole, we replaced the Newtonian potential by the Logarithmically
Modified Potential (LMP) proposed by
Mukhopadhyay~(\cite{Mukhopadhyay2003}).  This potential is well suited
to approximate the angular and epicyclic frequencies in accretion
disks around a rotating black hole. The expression of the radial
gravitational field derived from this potential is then given by~:
\begin{equation}
  \label{eq:GravitePseudoGR}
  g_\mathrm{r} = - \frac{G\,M_*}{r^2} \left[ 1 + R_{ms} \left( \frac{9}{20} \, 
      \frac{R_{ms}-1}{r} - \frac{3}{2r} \, \ln \frac{r}{(3r-R_{ms})^{2/9}} \right) \right]
\end{equation}
where~$R_{ms}$ is the last stable circular orbit~:
\begin{eqnarray}
  R_{ms} & = & 3+Z_2 \pm \sqrt{(3-Z_1)(3+Z_1+2Z_2)} \\
  Z_1 & = & 1 + (1-a^2)^{1/3} \, [ (1+a)^{1/3} + (1-a)^{1/3} ] \\
  Z_2 & = & \sqrt{3\,a^2 + Z_1^2}
\end{eqnarray}

The important feature of this potential is the vanishing of the radial
epicyclic frequency for a single particle having a circular orbit at
the innermost stable circular orbit (ISCO).

\begin{figure}[h]
  \begin{center}
    \includegraphics[draft=false,scale=0.6]{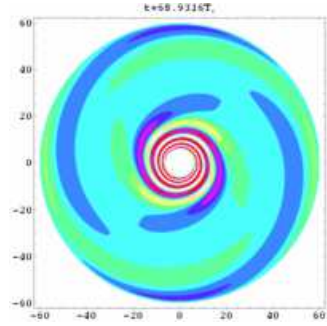}
    \caption{Final snapshot of the density perturbation in the accretion
      disk evolving in a perturbed pseudo-Schwarzschild potential
      (LMP). The disk extends from~$R_1=6.0$ to~$R_2=60.0$. No wave
      can propagate between the inner and outer Lindblad resonance.}
    \label{fig:DensDiscSchw} 
  \end{center}
\end{figure}

\begin{figure}[h]
  \begin{center}
    \includegraphics[draft=false,scale=0.8]{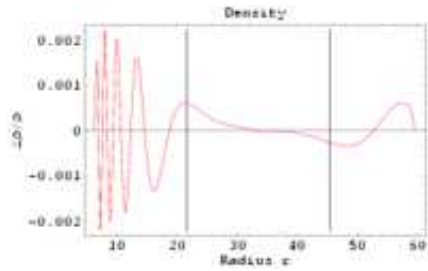}
    \caption{Same as Fig.~\ref{fig:DensDiscNewtSect} but for the
      pseudo-Schwarzschild geometry. The Lindblad resonances are
      located at~$r_\mathrm{L}^{in/out}=21.6/45.5$.}
    \label{fig:DensDiscSchwSect}
  \end{center}
\end{figure}

\begin{figure}[h]
  \begin{center}
    \includegraphics[draft=false,scale=0.8]{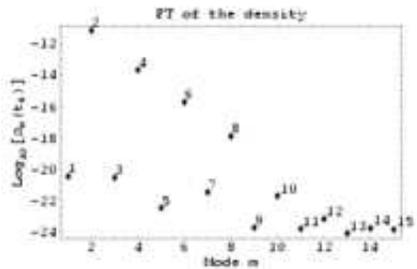} 
    \caption{Same as Fig.~\ref{fig:DensTFDiscNewt} but for the
      pseudo-Schwarzschild geometry. The Fourier coefficients follow a
      decaying geometric series.}
    \label{fig:DensTFDiscSchw} 
  \end{center}
\end{figure}

We use the same physical parameters as in the Newtonian case.  The
time evolution of the density perturbation in the disk is shown in
Fig.~\ref{fig:DensDiscSchw}. The Lindblad resonances are now located
at~$r_\mathrm{L}^{in/out}=21.6/45.5$ which differs slightly from the
previous simulation because the orbital velocity is no more the
Keplerian one but the pseudo-Newtonian one derived from
Eq.~(\ref{eq:GravitePseudoGR}). Here too, these locations are in
agreement with the linear analysis. After a few hundreds of orbital
revolutions, the disk settles down to a new quasi-stationary state,
very close to the one described by the linear analysis. The profile of
the density perturbation found by the numerical simulation is shown in
Fig.~\ref{fig:DensDiscSchwSect} to compare with the right plot of
Fig.~\ref{fig:SolForcee}. Note however that for the numerical
simulations, the radial epicyclic frequency derived from the LMP
Eq.~(\ref{eq:GravitePseudoGR}) differs slightly from the true one
given by Schwarzschild's solution Eq.~(\ref{eq:FreqRadial}). As can be
seen from Fig.~\ref{fig:DensTFDiscSchw}, the linear regime is still a
good approximation in this case, the dominant Fourier coefficient is
always~$m=2$.

To conclude this subsection, we have shown that the introduction of
the concept of ISCO in this last run does not change the qualitative
conclusions drawn by the Newtonian simulations. Its only effect is to
shift the location of the Lindblad resonance. This behavior was
expected from the linear analysis.

\subsection{Pseudo-Kerr potential}

The frame dragging effect induced by the star's rotation can also be
investigated by the pseudo-Newtonian potential described in the
previous section. Therefore we run a simulation in which the rotation
of the star is taken into account by the gravitational field described
by Eq.~(\ref{eq:GravitePseudoGR}).

To create a significant change in the orbital frequency, we choose a
spin parameter~$a_*=0.5$. Thus the disk inner boundary corresponding
to the marginally stable circular orbit is located at~$R_1=4.24$. The
outer computational domain is taken to be 10~times as large,
at~$R_2=42.4$.

\begin{figure}[h]
  \begin{center}
    \includegraphics[draft=false,scale=0.6]{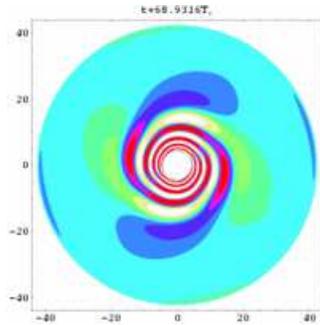}
    \caption{Final snapshot of the density perturbation in the accretion disk
      evolving in a perturbed pseudo-Kerr potential with~$a=0.5$. The
      disk extends from~$R_1=4.24$ to~$R_2=42.4$. The outer Lindblad
      resonance is not in the grid.}
    \label{fig:DensDiscKerr} 
  \end{center}
\end{figure}

\begin{figure}[h]
  \begin{center}
    \includegraphics[draft=false,scale=0.8]{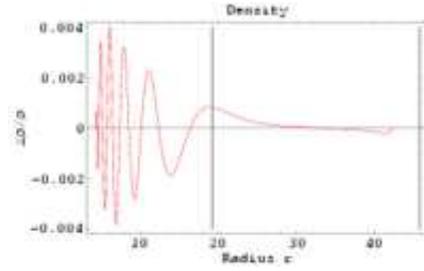}
    \caption{Same as Fig.~\ref{fig:DensDiscNewtSect} but for the pseudo-Kerr disk.
      The inner Lindblad resonance appears clearly
      at~$r_\mathrm{L}^{in}=19.2$ while the outer one
      at~$r_\mathrm{L}^{out}=45.8$ lies outside the computational
      grid.}
    \label{fig:DensDiscKerrSect} 
  \end{center}
\end{figure}

\begin{figure}[h]
  \begin{center}
    \includegraphics[draft=false,scale=0.8]{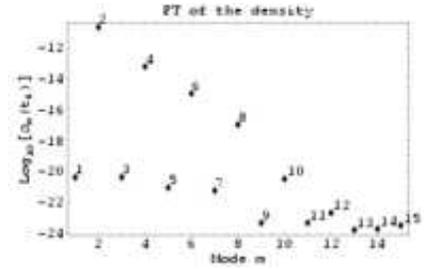}
    \caption{Same as Fig.~\ref{fig:DensTFDiscNewt} but for the pseudo-Kerr disk. 
      Again the Fourier coefficients decay like a geometric series.}
    \label{fig:DensTFDiscKerr} 
  \end{center}
\end{figure}

The inner Lindblad resonance is clearly identified
at~$r_\mathrm{L}^{in}=19.2$ as can be seen from
Fig.~\ref{fig:DensDiscKerr}. The outer Lindblad resonance
at~$r_\mathrm{L}^{out}=45.8$ lies outside the computational domain and
therefore does not show up in the plot,
Fig.~\ref{fig:DensDiscKerrSect}. The $m=2$ is the strongest mode
whereas the other odd modes decrease following a geometrical series as
confirmed by inspection of Fig.~\ref{fig:DensTFDiscKerr}.

Here again apart from the fact that the disk approaches closer to the
neutron star, the resonances behave in the same way as in the previous
sections. Thus we strongly believe that the QPO phenomenon has nothing
to do with any specific general relativistic effect. It just help to
tune the QPO frequencies to some given values which could not be
explained by a simple Newtonian gravitational potential.

\subsection{Retrograde disk}

In this run, we checked that the Lindblad resonances disappear for a
retrograde Newtonian disk. We rerun the simulation of
subsection~\ref{sec:ResNewt} by changing the sign of the spin of the
neutron star by~$\Omega_*=-0.0043311$. Thus the disk is rotating in
the opposite way compared to the star.

\begin{figure}[h]
  \begin{center}
    \includegraphics[draft=false,scale=0.6]{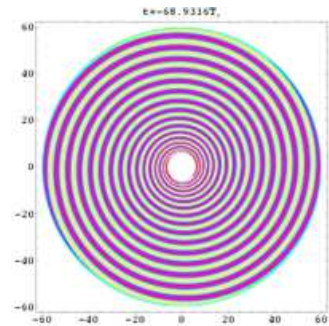}
    \caption{Final snapshot of the density perturbation in 
      the counterrotating accretion disk evolving in a perturbed
      Newtonian potential. Same values as in caption of
      Fig.~\ref{fig:DensDiscNewt} applies expected for the sign
      of~$\Omega_*$. A trailing spiral density wave of~$m=2$ is
      propagating in the whole disk.}
    \label{fig:DensRetroDiscNewt} 
  \end{center}
\end{figure}

\begin{figure}[h]
  \begin{center}
    \includegraphics[draft=false,scale=0.8]{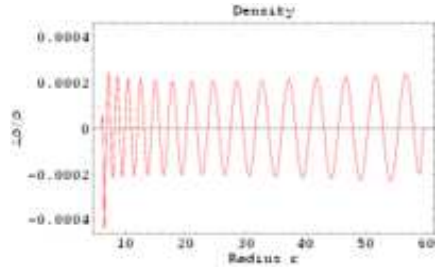}
    \caption{Same as Fig.~\ref{fig:DensDiscNewtSect} but for 
      the counterrotating disk. No Lindblad resonances are observed.}
    \label{fig:DensRetroDiscNewtSect} 
  \end{center}
\end{figure}

\begin{figure}[h]
  \begin{center}
    \includegraphics[draft=false,scale=0.8]{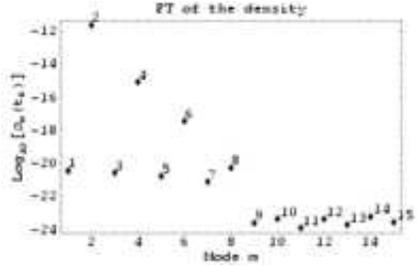}
    \caption{Same as Fig.~\ref{fig:DensTFDiscNewt} but for 
      the counterrotating disk.}
    \label{fig:DensTFRetroDiscNewt} 
  \end{center}
\end{figure}

As expected, no Lindblad resonance is observed in this run,
Fig.~\ref{fig:DensRetroDiscNewt}. The density profile perturbation is
depicted in Fig.~\ref{fig:DensRetroDiscNewtSect} agrees well with the
linear analysis, Fig.~\ref{fig:SolForceeContre}.  A long trailing wave
of mode~$m=2$ expands in the whole disk area. A quasi-stationary state
is reached very quickly. The Fourier spectrum behaves here again in
the same manner as in the other runs
Fig.~\ref{fig:DensTFRetroDiscNewt}.

\subsubsection{Wake solution in a protoplanetary disk}

\begin{figure}[htbp]
  \centering
  \includegraphics[scale=0.8]{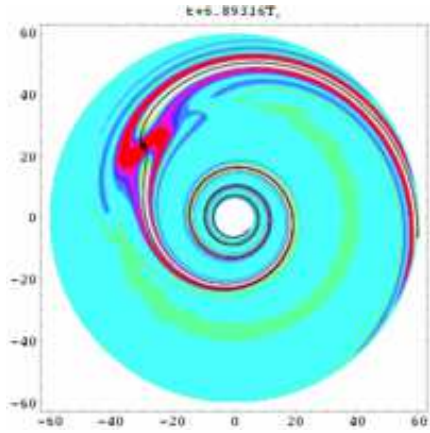} 
  \caption{One-armed spiral wake generated by the orbital motion of a planet 
    (or a small mass solid body) in the 2D accretion disk. The density
    perturbation~$\delta\rho/\rho$ is shown. The perturbing body (like
    a planet) is depicted as a black circle. The linear
    response~Eq.(\ref{eq:Wake}) is plotted as a solid line and
    overlaps well with the non-linear simulation.}
  \label{fig:Wake}
\end{figure}

\begin{figure}[h]
  \begin{center}
    \includegraphics[scale=0.8]{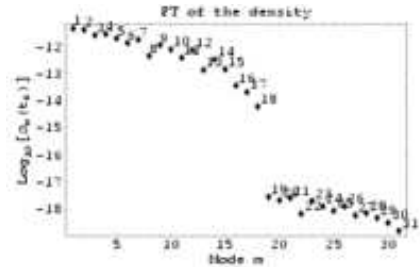} 
    \caption{Same as Fig.~\ref{fig:DensTFDiscNewt} but 
      for the wake solution.}
    \label{fig:WakeSect} 
  \end{center}
\end{figure}

We also performed a set of simulations in which a Keplerian disk
orbiting around a massive star~$M_*$ is perturbed by a planet of small
mass~$M_p\ll M_*$. In this situation, many azimuthal modes are excited
and propagate in the disk. The numerical algorithm is therefore
checked when many modes are excited at the same time. The solution is
represented by a one-armed spiral wake generated by the planet. In the
case of weak perturbation, Ogilvie \& Lubow~(\cite{Ogilvie2002}) have
shown that the linear response of the disk is obtained by constructive
interference between wave modes in the disk.  The approximate
analytical shape of the wake is given by~:
\begin{equation}
  \label{eq:Wake}
  \varphi = t \pm \frac{3}{2\,\varepsilon} ( \frac{r}{r_\mathrm{c}}^{3/2} - 
  \frac{3}{2} \, \ln \frac{r}{r_\mathrm{c}} -1)
\end{equation}
where $r_\mathrm{c}$ is the corotation radius and $t$ the time.  The
$+$ sign applies for the inner part of the disk~($r\le r_\mathrm{c}$)
while the $-$ sign applies for the outer part of the disk~($r\ge
r_\mathrm{c}$).  This result from the linear analysis is compared with
the full non-linear set of Euler equations.  An example is shown in
Fig.~\ref{fig:Wake}. The non-linear simulation agrees perfectly with
the linear solution given by~Eq.(\ref{eq:Wake}) (the resolution used
is~$256\times64$).  A one armed spiral wave is launched from the
rotating planet and propagates in the entire disk with a pattern speed
equal to the rotation rate of the perturber.  Several low azimuthal
modes are excited to a significant level as seen in
Fig.~\ref{fig:WakeSect}.  However, to avoid aliasing effect because of
the fast Fourier transform, we filtered out the high frequency
component for~$m\ge19$.  Moreover, the boundary condition imposed as
non reflective waves works well by damping the perturbations close to
the disk inner and outer edges.

\subsection{Realistic potential}

\begin{figure}[h]
  \begin{center}
    \includegraphics[draft=false,scale=0.6]{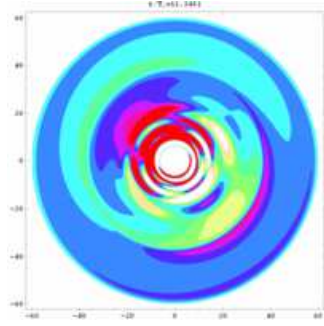}
    \caption{Final snapshot of the density perturbation in the accretion disk
      evolving in a perturbed Newtonian potential having many
      azimuthal modes~$m$. The disk extends from~$R_1=6$ to~$R_2=60$.}
    \label{fig:DensDiscBoiteux} 
  \end{center}
\end{figure}

\begin{figure}[h]
  \begin{center}
    \includegraphics[draft=false,scale=0.6]{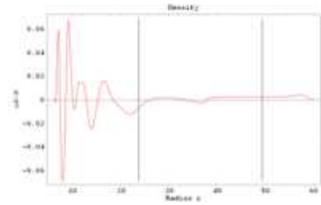}
    \caption{Same as Fig.~\ref{fig:DensDiscNewtSect} but 
      for a perturbation containing many modes.  The corotation
      resonance is shown by a vertical bar.}
    \label{fig:DensDiscBoiteuxSect} 
  \end{center}
\end{figure}

\begin{figure}[h]
  \begin{center}
    \includegraphics[draft=false,scale=0.8]{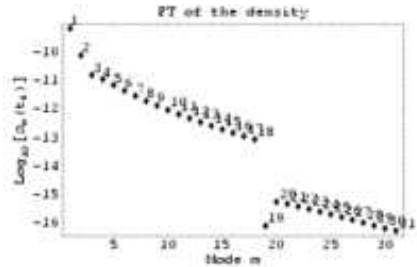}
    \caption{Same as Fig.~\ref{fig:DensTFDiscNewt} but for  
      a perturbation containing many modes. Only the lowest Fourier
      modes are excited to a significant level in accordance with the
      perturbed potential profile.}
    \label{fig:DensTFDiscBoiteux} 
  \end{center}
\end{figure}

In a last run, we went back to a multipolar gravitational perturbation
in the stellar disk as described in Sec.~\ref{sec:PotentielBoiteux}.
Because the gravitational field perturbation contains several Fourier
components given by the Laplace
coefficients~Eq.~(\ref{eq:CoeffLaplace}), many azimuthal modes are
excited as in the previous case of a protoplanetary disk.  For this
run, the numerical values are~$r_\mathrm{p}=5\,R_*, z_\mathrm{p}=0,
M_\mathrm{p} = 10^{-3}\,M_*$.  In order to keep a good numerical
accuracy even with a system containing many modes, we increased the
resolution by taking~$256\times64$. The final snapshot of the density
perturbation in the disk is shown in Fig.~\ref{fig:DensDiscBoiteux}.
The corotation radius is clearly identified at~$r=37.4$. The
fluctuation in density are relevant only in the innermost region of
the accretion disk where the tidal force is maximal. An inspection of
the Fig.~\ref{fig:DensDiscBoiteuxSect} confirms this remark. As seen
from Fig.~\ref{fig:DensTFDiscBoiteux}, the strongest mode is
associated with~$m=1$ that is the strongest exciting mode. Because the
system remains in a linear regime, its total response to the
perturbation is simply given by the sum of the response to each mode.
This explains the decrease in the perturbed fluctuation with the mode
number (see Fig.~\ref{fig:CoeffLaplace}). Here again, the desaliasing
process keeps the azimuthal modes $m\ge19$ close to zero (within the
numerical accuracy).

\section{DISCUSSION}
\label{sec:Discussion}

The simulations presented in this paper are preliminary results mainly
in order to check the numerical algorithm and to show that
pseudo-spectral method are well suited to study accretion flow as was
already done by Godon~(\cite{Godon1997}). For smooth flows, this
method shows evanescent error (exponential decrease with respect to
the number of points) and good accuracy is achieved with relatively
small number of discretization points. 

The warping of the disk is an important processes to account for QPOs
because the vertical parametric resonance locations are interesting
sites to generate large amplitude in the fluctuation of the luminosity
of the accretion disk.  A full three dimensional linear analysis is
therefore necessary to track the propagation properties of the waves
in a polytropic disk.  Moreover, non-linear effect are also important
because in the case of magnetized accretion disk, the rotating
asymmetric perturbed magnetic field is of the same order of magnitude
as the background field itself. Full non-linear three dimensional
numerical simulations are therefore required. The numerical code
designed here can performed this task with a reasonable computational
time by using only little discretization points without losing
precision.

The study presented in this paper can be extended to the case of a
viscous flow in the disk leading therefore to a ``real accretion
disk''. The viscosity will set a lower limit on the smallest scale
that the density perturbation can reach. Moreover, when adding a
damping term in the Mathieu equation (\ref{eq:Mathieu}) the threshold
for the parametric resonance to grow will be proportional to this
viscosity. We therefore expect strong oscillations only in cases where
the amplitude of the perturbation is sufficiently high.  In addition,
the lost of energy by radiation damps these oscillations for which the
kinetic energy is converted into photon emission. Finally, because the
radial flux implied by the accretion process will reduce the time
available for a fluid element to enter into resonance at a given
radius, deviation from equilibrium will become smaller compared to a
non accreting situation.

Another point not discussed in this work is the pressure induced by
the radiation in the vicinity of the inner part of the accretion disk.
As for the gaseous pressure, it will shift the resonance conditions
derived for a single particle. A significant increase in the accretion
flux induces an increase in the pressure radiation to the same order
of magnitude. How the disk will react remains an open question.
However it provides a correlation between the accretion rate and the
orbital frequency at resonance because the resonance conditions
Eq.~(\ref{eq:ResCorot})-(\ref{eq:ResForcage})-(\ref{eq:ResPara}) will
then depend on the total pressure in the disk (gaseous + radiation).

The origin of the rotating asymmetric gravitational field for neutron
stars or white dwarfs could be explained in severals ways.  First, an
inhomogeneous interior structure would naturally lead to the kind of
potential introduced in Sec.~\ref{sec:PotentielBoiteux}.  A second
possibility is the fast rotation of the star. It leads to strong
centrifugal forces deforming the star's spherical shape into a
Maclaurin spheroid. Thirdly, an aspherical star possessing a
precession motion should also generate a significant distorted
gravitational potential. Anisotropic magnetic stress in the stellar
interior exerts also a deformation on the stellar surface which can be
inferred by observation of the atmosphere. An application to white
dwarfs is given by Fendt \& Dravins~(\cite{Fendt2000}) and for neutron
star by Bonazzola \& Gourgoulhon~(\cite{Bonazzola1996}).
  
For the black hole candidates, the reason for an asymmetry in the
spacetime manifold is less evident. When accreting matter, the black
hole must get ride of its non-stationary state (deviation from the
Kerr-Newman geometry) because of the no-hair theorem. In order to get
back to its stationary state, it must emit gravitational waves
described in the general relativity framework. This replaces the
rotating asymmetric Newtonian potential. Like helioseismology,
perturbation of the spacetime around black holes gives some insight
into their properties.  However, unlike asteroseismology, normal mode
oscillations are unavoidably associated to gravitational wave
emission. They are therefore called quasi-normal modes of the black
hole because of their damping (Kokkotas \& Schmidt
\cite{Kokkotas1999}).  If the QPOs observed in the black hole
candidates could be associated with the gravitational wave emission
and thus with their quasi-normal modes, we would have a new tool to
explore their properties as was the case for helioseismology, a kind
of ``holoseismology''. This idea is discussed in P\'etri
(\cite{Petri2005d}).
  
The discovery of high coherence in the kHz-QPOs of some systems puts
strong constraints on the models. Clumps of matter cannot account for
the high quality factor $Q>100$ (Barret et al.  \cite{Barret2005a}).
The coherence time involved by this picture is to long. We believe
that this almost sinusoidal motion can only be imprinted by an
external sinusoidal force as for instance a rotating asymmetric field
would do. Moreover, sometimes, a sudden drop in this coherence is
observed as was happened in 4U1636-536. Like the discrimination
between slow and fast rotators~(P\'etri \cite{Petri2005b},
\cite{Petri2005c}), it is interpreted as another manifestation of the
ISCO (Barret et al. \cite{Barret2005b}).

\section{CONCLUSION}
\label{sec:Conclusion}

In this paper, we have explored the consequences of a weak rotating
asymmetric gravitational potential perturbation on the evolution of a
thin accretion disk initially in a stationary axisymmetric state. We
have shown that, when one mode is excited, the disk resonates at some
radii where the resonance conditions are satisfied and reach a new
quasi-stationary state in which some small scale perturbations in the
density emanate starting from the inner and outer Lindblad resonance.
The non wavelike disturbance rotates at the star's spin while the free
wave solutions are irrelevant because of damping process. Only the
driven resonance can be maintained and account for QPOs having very
high quality factors. For more general gravitational perturbation
potentials, the response of the disk is the some of individual modes
as long as the system remains in the linear regime. We gave a example
of such a set of resonances in the last part. The physical processes
at hand does not require any general relativistic effect. Indeed, the
resonances behave identical in the Newtonian as well as in the
pseudo-Kerr field.  However, in order to get a detailed precise
quantitative idea of the properties of free wave solutions and non
wavelike disturbances around neutron star, we need a consistent full
general relativistic description of the star-disk system. This is left
for future work.

The possible warping of the disk, not discussed here, needs a full 3D
analysis and simulations. We then expect some low frequency QPOs
related to the kHz QPOs in a manner which has still to be determined.

We emphasize that the work presented here is a first step to a new
model for the QPOs in LMXBs. A strongly non linear regime is also
expected when the gravity perturbation is strong enough. The
parametric resonance not developed in the simulations presented in
this paper will then be excited. This is left for future work.

To conclude, to date we know about 20~LMXBs containing a neutron star
and all of them show kHz-QPOs. We believe that these QPOs could be
explained by a mechanism similar to those exposed here. We need only
to replace the gravitational perturbation by a magnetic one as
described in paper~I. However, in an accreting system in which the
neutron star is an oblique rotator, we expect a perturbation in the
magnetic field to the same order of magnitude than the unperturbed
one. Therefore, the linear analysis developed in this series of two
papers has to be extended to oscillations having non negligible
amplitude compared to the stationary state. We also expect the
parametric resonance to become the strongest resonance in the disk.
Indeed, as shown in previous work in the single particle approximation
and in the MHD case (P\'etri \cite{Petri2005b}, \cite{Petri2005c}),
the twin peak ratio of about 3/2 for the kHz-QPOs is naturally
explained as well as their difference being either $\Delta\nu=\nu_*$
for slow rotator ($\nu_*\le300$~Hz) or $\Delta\nu=\nu_*/2$ for fast
rotator ($\nu_*\ge300$~Hz).

Finally, what would happen if we add a gravitational perturbation to a
magnetic one? In the thin and weakly magnetised accretion disk, the
linear analysis performed separately in the hydrodynamical as well as
in the MHD case remains true when combined together. We expect again
the same resonances to occur at the same locations.  This is the
strength of this model because it encompasses in an unique picture the
white dwarf (though to be mainly in the hydrodynamical regime) and
neutron star (though to be magnetised) accreting systems, explaining
the 3:2 ratio whatever the nature of the compact object. In has also
successfully been extended to accreting black holes by replacing these
asymmetries by gravitational wave emission (P\'etri
\cite{Petri2005d}). The predicted 3:2 ratio and some lower frequency
QPOs perfectly match the observations from the microquasar
GRS1915+105.

\begin{acknowledgements}
  This research was carried out in a FOM projectruimte on
  `Magnetoseismology of accretion disks', a collaborative project
  between R. Keppens (FOM Institute Rijnhuizen, Nieuwegein) and N.
  Langer (Astronomical Institute Utrecht). This work is part of the
  research programme of the `Stichting voor Fundamenteel Onderzoek der
  Materie (FOM)', which is financially supported by the `Nederlandse
  Organisatie voor Wetenschappelijk Onderzoek (NWO)'.
  
  This work was also supported by a grant from the G.I.F., the
  German-Israeli Foundation for Scientific Research and Development.
\end{acknowledgements}


%
%

\appendix


\section{Derivation of the thin disk eigenvalue problem}
\label{sec:AppDerSystProp}

In this appendix, we derive the eigenvalue problem satisfied by the
radial Lagrangian displacement~$\vec{\xi}$ for a thin accretion disk
for which~$(H/R) \approx (c_\mathrm{s}/R\,\Omega) \ll 1$. We focus only on the fluid
motion in the plane of the disk, no warp is taken into account, and so
$\xi_\mathrm{z}=0$.  Projecting Eq.~(\ref{eq:PDEXi}) on the radial and
azimuthal axis, we get the evolution equations for the 2D Lagrangian
displacement~$(\xi_\mathrm{r},\xi_\varphi)$ as~:
\begin{eqnarray}
  \label{eq:AppPDEXiR2D}
  & & \rho \, \frac{\partial^2\xi_\mathrm{r}}{\partial t^2} + 2 \, \rho \, \Omega \, \left(
    \frac{\partial^2\xi_\mathrm{r}}{\partial\varphi\partial t} - 
    \frac{\partial\xi_\varphi}{\partial t} \right)
  - \frac{\partial}{\partial r} \, \left( \gamma \, p \, \div\vec{\xi} + \vec{\xi} \cdot 
    \vec{\nabla} p \right) + (g_\mathrm{r} + r\,\Omega^2 )\,\div(\rho\,\vec{\xi}) + \nonumber \\ 
  & & \rho \, \xi_\mathrm{r} \, \frac{d}{dr} ( r \, \Omega^2) + 
  \rho \, \Omega^2 \left( \frac{\partial^2\xi_\mathrm{r}}{\partial\varphi^2}
    - 2\,\frac{\partial\xi_\varphi}{\partial\varphi} - \xi_\mathrm{r} \right) = \rho \, \delta g_\mathrm{r} \\
  \label{eq:AppPDEXiP2D}
  & & \rho \, \frac{\partial^2\xi_\varphi}{\partial t^2} + 2 \, \rho \, \Omega \, \left(
    \frac{\partial^2\xi_\varphi}{\partial\varphi\partial t} + 
    \frac{\partial\xi_\mathrm{r}}{\partial t} \right)
  - \frac{\partial}{r\,\partial\varphi} \, \left( \gamma \, p \, \div\vec{\xi} + 
    \vec{\xi} \cdot \vec{\nabla} p \right) + \nonumber \\
  & & \rho \, \Omega^2 \left( \frac{\partial^2\xi_\varphi}{\partial\varphi^2}
    + 2\,\frac{\partial\xi_\mathrm{r}}{\partial\varphi} \right) + g_\varphi \, ( \rho\,\div\vec{\xi} +
  \vec{\xi} \cdot \vec{\nabla}\rho ) = \rho \, \delta g_\varphi
\end{eqnarray}

These are the two non-homogeneous equation for the Lagrangian
displacement in the disk. The perturbation in the gravitational field
gives rise to a driving force responsible for the non-homogeneous
part.  The solutions of these equations consist of free wave solutions
and non-wavelike perturbations. In both cases, we are looking for
solutions expressed as a plane wave in the~$(\varphi,t)$ coordinates~:
\begin{equation}
  \label{eq:AppOndePlane}
  X(r,\varphi,t) = X(r) \, e^{i(m\varphi-\sigma\,t)}
\end{equation}
For the non-wavelike solution, the eigenfrequency is imposed by the
rotating star and is given by~$\sigma = m\,\Omega_*$. Putting the
development~Eq.~(\ref{eq:AppOndePlane}) into the
system~(\ref{eq:AppPDEXiR2D})-(\ref{eq:AppPDEXiP2D}), we obtain~:
\begin{eqnarray}
  \label{eq:AppPDEXiR2D2}
  (\omega^2 - r\,\frac{d}{dr}(\Omega_\mathrm{k}^2) + \Omega^2 - \Omega_\mathrm{k}^2) \, \xi_\mathrm{r}
  - 2\,i\,\Omega_\mathrm{k} \, \omega \, \xi_\varphi & + & \nonumber \\
  c_\mathrm{s}^2 \, \frac{\partial}{\partial r} (\div\vec{\xi})
  + \frac{1}{\rho} \, \frac{\partial p}{\partial r} \, \left[ (\gamma-1) \, \div\vec{\xi} 
    + \frac{\partial\xi_\mathrm{r}}{\partial r} \right] & = & - \delta g_\mathrm{r} \\
  \label{eq:AppPDEXiP2D2}
  \omega^2 \, \xi_\varphi + 2\,i\,\Omega_\mathrm{k} \, \omega \, \xi_\mathrm{r} + \frac{i\,m}{r} \, 
  \left( c_\mathrm{s}^2 \, \div\vec{\xi} + \frac{1}{\rho} \, 
    \frac{\partial p}{\partial r} \, \xi_\mathrm{r} \right) & = & - \delta g_\varphi
\end{eqnarray}
We have introduce the following notation~:
\begin{itemize}
\item the Keplerian angular velocity~:
  \begin{equation}
    \label{eq:AppFreqKepl}
    \Omega_\mathrm{k} = \sqrt{\frac{G\,M_*}{r^3}}
  \end{equation}
\item the Doppler shifted eigenfrequency~:
  \begin{equation}
    \label{eq:AppValPropDoppler}
    \omega = \sigma - m\,\Omega
  \end{equation}
\item the divergence of the complex vector~$\vec{\xi}(\vec{r})$~:
  \begin{equation}
    \label{eq:AppDivXi}
    \div\vec{\xi} = \frac{1}{r} \, \frac{\partial}{\partial r} (r\,\xi_\mathrm{r}) + 
     \frac{i\,m}{r} \, \xi_\varphi 
  \end{equation}
\end{itemize}
These are the generalization of the equations obtained by Nowak \&
Wagoner~(\cite{Nowak1991}) in case of an external force acting on the
disk.  We can then solve Eq.~(\ref{eq:AppPDEXiP2D2}) for the
variable~$\xi_\varphi$ by~:
\begin{equation}
  \label{eq:AppXiPhi}
  \xi_\varphi = - \frac{1}{\omega_*^2} \, \left[ \delta g_\varphi + i \,
    \left(  2\,\Omega\,\omega + \frac{m}{r} \, \frac{1}{\rho} \, 
      \frac{\partial p}{\partial r} \right) \, \xi_\mathrm{r} + i \, \frac{m\,c_\mathrm{s}^2}{r^2} \,
    \frac{\partial}{\partial r} (r\,\xi_\mathrm{r}) \right]
\end{equation}
with~$\omega_*^2 = \omega^2 - \frac{m^2}{r^2} \, c_\mathrm{s}^2$.

Substituting in the Eq.~(\ref{eq:AppPDEXiR2D2}), the radial displacement
satisfies a second order linear differential equation~:
\begin{eqnarray}
  \label{eq:AppXiR}
  c_\mathrm{s}^2 \, \left[ 1 + \frac{m^2\,c_\mathrm{s}^2}{r^2\,\omega_*^2} \right] \,
  \frac{\partial^2\xi_\mathrm{r}}{\partial r^2} + 
  \frac{c_\mathrm{s}^2}{r} \, \left[ \frac{\partial \ln\,(rp)}{\partial\ln\,r} + 
    \frac{m^2\,c_\mathrm{s}^2}{r} \, \frac{\partial}{\partial r}(\omega_*^{-2}) - 
    \frac{m^2\,c_\mathrm{s}^2}{\omega_*^2\,r^2} 
    + \frac{m^2\,c_\mathrm{s}^2}{\omega_*^2\,r} \, \frac{\partial\ln p}{\partial r} +
    2 \, \frac{m^2\,c_\mathrm{s}}{\omega_*^2\,r} \, \frac{\partial c_\mathrm{s}}{\partial r} \right] \,
  \frac{\partial\xi_\mathrm{r}}{\partial r} + \nonumber \\
  \left[ \omega^2 - \kappa^2 + 4 \, \Omega^2\, \left( 1 - \frac{\omega^2}{\omega_*^2} \right) + 
    + r \, \frac{\partial}{\partial r}(\Omega^2-\Omega_0^2) + \frac{c_\mathrm{s}^2}{r} \, 
    ( \frac{\partial\ln p}{\partial r} - \frac{1}{r} ) +
  \right. \nonumber \\ \left.
    \frac{m\,c_\mathrm{s}^2}{r} \, \left( 2\,\Omega\,\omega + \frac{m}{r} \, \left( 
        \frac{c_\mathrm{s}^2}{\gamma} \, \frac{\partial\ln p}{\partial r} + 
        \frac{c_\mathrm{s}^2}{r} \right) \right) \, \frac{\partial}{\partial r}(\omega_*^{-2}) +
    \frac{\gamma-1}{\gamma\,r} \, c_\mathrm{s}^2 \, \frac{\partial\ln p}{\partial r} \, 
    \frac{m}{\omega_*^2} \, \left( 2\,\Omega\,\omega + \frac{m}{r} \, \left( 
        \frac{c_\mathrm{s}^2}{\gamma} \, \frac{\partial\ln p}{\partial r} + 
        \frac{c_\mathrm{s}^2}{r} \right) \right) +   \right. \nonumber \\ \left.
    \frac{m}{\omega_*^2} \left( m\,c_\mathrm{s}^2 \frac{\partial}{\partial r} \frac{c_\mathrm{s}^2}{r^2} + 
      c_\mathrm{s}^2 \, \frac{\partial}{\partial r} 
      \left( \frac{2\,\Omega\,\omega}{r} + \frac{m\,c_\mathrm{s}^2}{\gamma\,r^2} \, 
        \frac{\partial\ln p}{\partial r} \right) - \frac{2\,\Omega\,\omega}{r} \, \left( 
        \frac{c_\mathrm{s}^2}{\gamma} \, \frac{\partial\ln p}{\partial r} + 
        \frac{c_\mathrm{s}^2}{r} \right) \right) \right] \, \xi_\mathrm{r} \nonumber \\
  = -\delta g_\mathrm{r} - i \left[ \left( 2 \, \frac{\Omega\,\omega}{\omega_*^2} -
      \frac{\gamma-1}{r\,\gamma} \, \frac{m\,c_\mathrm{s}^2}{\omega_*^2} \, 
      \frac{\partial\ln p}{\partial r} + \frac{m\,c_\mathrm{s}^2}{r^2\,\omega_*^2} - 
      \frac{m\,c_\mathrm{s}^2}{r} \, \frac{\partial}{\partial r}
      \left( \omega_*^{-2} \right) \right) \, \delta g_\varphi - 
    \frac{m\,c_\mathrm{s}^2}{r\,\omega_*^2} \, \frac{\partial\delta g_\varphi}{\partial r} \right] 
  \nonumber \\
& & 
\end{eqnarray}

This equation can be drastically simplified in the case of a thin disk
where the sound speed can be considered as a small quantity.  However,
in order to include properly the corotation resonance, we keep the
leading terms close to the corotation radius and we have~:
\begin{eqnarray}
  \label{eq:AppXiRCorot}
  c_\mathrm{s}^2 \, \left[ 1 + \frac{m^2\,c_\mathrm{s}^2}{r^2\,\omega_*^2} \right] \,
  \frac{\partial^2\xi_\mathrm{r}}{\partial r^2} + \frac{c_\mathrm{s}^2}{r} \, \left[ 
    \frac{\partial \ln\,(rp)}{\partial\ln\,r} + \frac{m^2\,c_\mathrm{s}^2}{r} \, 
    \frac{\partial}{\partial r}(\omega_*^{-2}) \right] \,
  \frac{\partial\xi_\mathrm{r}}{\partial r} + \nonumber \\
  \left[ \omega^2 - \kappa^2 + 4 \, \Omega^2\, 
    \left( 1 - \frac{\omega^2}{\omega_*^2} \right) + 
    \frac{m\,c_\mathrm{s}^2}{r} \, 2\,\Omega\,\omega \, 
    \frac{\partial}{\partial r}(\omega_*^{-2}) \right] \, \xi_\mathrm{r} \nonumber \\
  = - \delta g_\mathrm{r} - i \left[ 2 \, \frac{\Omega\,\omega}{\omega_*^2} - 
    \frac{m\,c_\mathrm{s}^2}{r} \, \frac{\partial}{\partial r}
    \left( \omega_*^{-2} \right) \right] \, \delta g_\varphi 
\end{eqnarray}

Furthermore, when far from the corotation, terms including the sound
speed~$c_\mathrm{s}^2$ can be neglected compared to other ones, and
Eq.~(\ref{eq:AppXiRCorot}) reduces to~:
\begin{equation}
  \label{eq:AppXiRSimp}
  c_\mathrm{s}^2 \, \left[ \frac{\partial^2\xi_\mathrm{r}}{\partial r^2} +
    \frac{\partial \ln\,(rp)}{\partial\,r} \,
    \frac{\partial\xi_\mathrm{r}}{\partial r} \right] + ( \omega^2 - \kappa^2 ) \, \xi_\mathrm{r} =
  - \delta g_\mathrm{r} - 2 \, i \, \frac{\Omega}{\omega} \, \delta g_\varphi
\end{equation}

Finally we introduce the new unknown~$\psi=\sqrt{r\,p}\,\xi_\mathrm{r}$ to get
to the same order of approximation the following Schr\"odinger type
equation~:
\begin{equation}
  \label{eq:AppPsiSimp}
  \psi''(r) + V(r) \, \psi(r) = F(r)
\end{equation}
The potential is given by~:
\begin{equation}
  \label{eq:AppPotentiel}
  V(r) = \frac{\omega^2 - \kappa^2}{c_\mathrm{s}^2}
\end{equation}
and the source term by~:
\begin{equation}
  \label{eq:AppSrc}
  F(r) = - \left( \delta g_\mathrm{r} + 2 \, i \, \frac{\Omega}{\omega} \, 
    \delta g_\varphi \right) \, \frac{\sqrt{r\,p}}{c_\mathrm{s}^2}
\end{equation}
Eq.~(\ref{eq:AppPsiSimp}) is the fundamental equation we have to solve
to find the solution to our problem far from the corotation resonance.
We refer the reader to Appendix~B of paper~I where we develop an
analytical method for approximate solutions to this Schr\"odinger
equation.

\end{document}